\documentclass[ALICE,manyauthors]{cernphprep}
\usepackage[comma,square,numbers,sort&compress]{natbib}
\usepackage{hyperref}
\usepackage{lineno}
\usepackage{xspace}
\usepackage{multicol}
\usepackage{multirow}
\usepackage{booktabs}
\usepackage{threeparttable}
\usepackage{tabularx}
\usepackage{footnote}
\usepackage{blindtext}
\usepackage[T1]{fontenc}
\usepackage{orcidlink}

\begin{document}
%

\newcommand{\pp}           {pp\xspace}
\newcommand{\ppbar}        {\mbox{$\mathrm {p\overline{p}}$}\xspace}
\newcommand{\XeXe}         {\mbox{Xe--Xe}\xspace}
\newcommand{\PbPb}         {\mbox{Pb--Pb}\xspace}
\newcommand{\pA}           {\mbox{pA}\xspace}
\newcommand{\pPb}          {\mbox{p--Pb}\xspace}
\newcommand{\AuAu}         {\mbox{Au--Au}\xspace}
\newcommand{\dAu}          {\mbox{d--Au}\xspace}

\newcommand{\s}            {\ensuremath{\sqrt{s}}\xspace}
\newcommand{\snn}          {\ensuremath{\sqrt{s_{\mathrm{NN}}}}\xspace}
\newcommand{\pt}           {\ensuremath{p_{\rm T}}\xspace}
\newcommand{\meanpt}       {$\langle p_{\mathrm{T}}\rangle$\xspace}
\newcommand{\ycms}         {\ensuremath{y_{\rm CMS}}\xspace}
\newcommand{\ylab}         {\ensuremath{y_{\rm lab}}\xspace}
\newcommand{\etarange}[1]  {\mbox{$\left | \eta \right |~<~#1$}}
\newcommand{\yrange}[1]    {\mbox{$\left | y \right |~<~#1$}}
\newcommand{\dndy}         {\ensuremath{\mathrm{d}N_\mathrm{ch}/\mathrm{d}y}\xspace}
\newcommand{\dndeta}       {\ensuremath{\mathrm{d}N_\mathrm{ch}/\mathrm{d}\eta}\xspace}
\newcommand{\avdndeta}     {\ensuremath{\langle\dndeta\rangle}\xspace}
\newcommand{\dNdy}         {\ensuremath{\mathrm{d}N_\mathrm{ch}/\mathrm{d}y}\xspace}
\newcommand{\Npart}        {\ensuremath{N_\mathrm{part}}\xspace}
\newcommand{\Ncoll}        {\ensuremath{N_\mathrm{coll}}\xspace}
\newcommand{\dEdx}         {\ensuremath{\textrm{d}E/\textrm{d}x}\xspace}
\newcommand{\RpPb}         {\ensuremath{R_{\rm pPb}}\xspace}

\newcommand{\nineH}        {$\sqrt{s}~=~0.9$~Te\kern-.1emV\xspace}
\newcommand{\seven}        {$\sqrt{s}~=~7$~Te\kern-.1emV\xspace}
\newcommand{\twoH}         {$\sqrt{s}~=~0.2$~Te\kern-.1emV\xspace}
\newcommand{\twosevensix}  {$\sqrt{s}~=~2.76$~Te\kern-.1emV\xspace}
\newcommand{\five}         {$\sqrt{s}~=~5.02$~Te\kern-.1emV\xspace}
\newcommand{\twosevensixnn}{$\sqrt{s_{\mathrm{NN}}}~=~2.76$~Te\kern-.1emV\xspace}
\newcommand{\fivenn}       {$\sqrt{s_{\mathrm{NN}}}~=~5.02$~Te\kern-.1emV\xspace}
\newcommand{\LT}           {L{\'e}vy-Tsallis\xspace}
\newcommand{\GeVc}         {Ge\kern-.1emV/$c$\xspace}
\newcommand{\MeVc}         {Me\kern-.1emV/$c$\xspace}
\newcommand{\TeV}          {Te\kern-.1emV\xspace}
\newcommand{\GeV}          {Ge\kern-.1emV\xspace}
\newcommand{\MeV}          {Me\kern-.1emV\xspace}
\newcommand{\GeVmass}      {Ge\kern-.2emV/$c^2$\xspace}
\newcommand{\MeVmass}      {Me\kern-.2emV/$c^2$\xspace}
\newcommand{\lumi}         {\ensuremath{\mathcal{L}}\xspace}

\newcommand{\ITS}          {\rm{ITS}\xspace}
\newcommand{\TOF}          {\rm{TOF}\xspace}
\newcommand{\ZDC}          {\rm{ZDC}\xspace}
\newcommand{\ZDCs}         {\rm{ZDCs}\xspace}
\newcommand{\ZNA}          {\rm{ZNA}\xspace}
\newcommand{\ZNC}          {\rm{ZNC}\xspace}
\newcommand{\SPD}          {\rm{SPD}\xspace}
\newcommand{\SDD}          {\rm{SDD}\xspace}
\newcommand{\SSD}          {\rm{SSD}\xspace}
\newcommand{\TPC}          {\rm{TPC}\xspace}
\newcommand{\TRD}          {\rm{TRD}\xspace}
\newcommand{\VZERO}        {\rm{V0}\xspace}
\newcommand{\VZEROA}       {\rm{V0A}\xspace}
\newcommand{\VZEROC}       {\rm{V0C}\xspace}
\newcommand{\Vdecay} 	   {\ensuremath{V^{0}}\xspace}

\newcommand{\ee}           {\ensuremath{e^{+}e^{-}}} 
\newcommand{\pip}          {\ensuremath{\pi^{+}}\xspace}
\newcommand{\pim}          {\ensuremath{\pi^{-}}\xspace}
\newcommand{\kap}          {\ensuremath{\rm{K}^{+}}\xspace}
\newcommand{\kam}          {\ensuremath{\rm{K}^{-}}\xspace}
\newcommand{\pbar}         {\ensuremath{\rm\overline{p}}\xspace}
\newcommand{\kzero}        {\ensuremath{{\rm K}^{0}_{\rm{S}}}\xspace}
\newcommand{\lmb}          {\ensuremath{\Lambda}\xspace}
\newcommand{\almb}         {\ensuremath{\overline{\Lambda}}\xspace}
\newcommand{\Om}           {\ensuremath{\Omega^-}\xspace}
\newcommand{\Mo}           {\ensuremath{\overline{\Omega}^+}\xspace}
\newcommand{\X}            {\ensuremath{\Xi^-}\xspace}
\newcommand{\Ix}           {\ensuremath{\overline{\Xi}^+}\xspace}
\newcommand{\Xis}          {\ensuremath{\Xi^{\pm}}\xspace}
\newcommand{\Oms}          {\ensuremath{\Omega^{\pm}}\xspace}
\newcommand{\degree}       {\ensuremath{^{\rm o}}\xspace}

\begin{titlepage}
\PHyear{2023}       
\PHnumber{264}      
\PHdate{17 November}  

\title{Multiplicity dependence of charged-particle intra-jet properties in \pp collisions at $\sqrt{\textbf{\textit{s}}}$ = 13\,TeV}

\ShortTitle{Jet properties in pp collisions at $\sqrt{s}$ = 13\,TeV}   

\Collaboration{ALICE Collaboration\thanks{See Appendix~\ref{app:collab} for the list of collaboration members}}
\ShortAuthor{ALICE Collaboration} 

\begin{abstract}
  The first measurement of the multiplicity dependence of intra-jet properties of leading charged-particle jets in proton--proton (pp) collisions is reported. The mean charged-particle multiplicity and jet fragmentation distributions are measured in minimum-bias and high-multiplicity pp collisions at center-of-mass energy $\sqrt{s}$ = 13 TeV using the ALICE detector. Jets are reconstructed from charged particles produced in the midrapidity region ($|\eta| < 0.9$) using the sequential recombination anti-$k_{\rm T}$ algorithm with jet resolution parameters $R$ = 0.2, 0.3, and 0.4 for the transverse momentum (\pt) interval 5--110 GeV/$c$. The high-multiplicity events are selected by the forward V0 scintillator detectors. The mean charged-particle multiplicity inside the leading jet cone rises monotonically with increasing jet \pt in qualitative agreement with previous measurements at lower energies. The distributions of jet fragmentation function variables $z^{\rm ch}$ and $\xi^{\rm ch}$ are measured for different jet-\pt intervals. Jet-\pt independent fragmentation of leading jets is observed for wider jets except at high- and low-$z^{\rm ch}$ values. The observed ``hump-backed plateau'' structure in the $\xi^{\rm ch}$ distribution indicates suppression of low-\pt particles. In high-multiplicity events, an enhancement of the fragmentation probability of low-$z^{\rm ch}$ particles accompanied by a suppression of  high-$z^{\rm ch}$ particles is observed compared to minimum-bias events. This behavior becomes more prominent for low-\pt jets with larger jet radius. The results are compared with predictions of QCD-inspired event generators, PYTHIA\,8 with Monash 2013 tune and EPOS LHC. It is found that PYTHIA\,8 qualitatively reproduces the jet modification in high-multiplicity events except at high jet \pt. These measurements provide important constraints to models of jet fragmentation.
\end{abstract}
\end{titlepage}

\setcounter{page}{2} 

\section{Introduction}

Hadronic and nuclear collisions at ultra-relativistic energies are the subject of intense research in the field of high-energy physics, as they enable the study of the fundamental constituents of matter and the forces that govern their interactions. The energy density achieved in the laboratory by colliding high-energy nucleus beams is sufficient to allow the confined hadronic matter to be transformed into a hot and dense state of quantum chromodynamics (QCD)~\cite{QCD1}  matter where partons are no longer confined into hadrons, known as quark--gluon plasma (QGP)~\cite{QGP1,QGP2,QGP3,ALICEQCD}. Several experiments at RHIC and the LHC are being performed to study the physics of this strongly-interacting QCD matter. Various experimental signatures have been observed in heavy-ion (A--A) collisions in favor of the formation of the QGP medium. Jet quenching~\cite{jetquench1,jetquench2, JETquenchBurke}, which particularly manifests as in-medium energy loss of energetic partons~\cite{JETSCAPEenergyloss,STARenergyloss,Wangenergyloss,Zappenergyloss}, is one of the most important signatures among them.

Jets are cascades of energetic hadrons that result from the fragmentation of hard-scattered (i.e. produced in processes with large squared momentum transfer $Q^{\rm 2}$) quarks and gluons in high-energy collisions. In proton--proton (pp) collisions, measurements of jet production provide a test bench of perturbative calculations and help to study non-perturbative effects in QCD~\cite{ALICExsec1,ALICExsec2,ALICExsec3,ALICExsec4,ATLASxsec1}. In addition, measurements in pp collisions also provide the baseline for similar measurements in A--A collisions. The number of reconstructed jets is found to be suppressed, and jet properties are also modified with respect to those in pp collisions due to the presence of hot and dense QCD matter in A--A collisions. This phenomenon is known as jet quenching in heavy-ion collisions. In particular, highly energetic jets, while passing through the QGP medium, lose energy via elastic scatterings and medium-induced gluon radiations. 

Interestingly, recent measurements of high-multiplicity \pp and \pPb (proton-lead) collisions show ample signatures conventionally associated with the QGP formation in heavy-ion collisions. These observations triggered an immense research interest to look for the onset of QGP-like effects in high-energy small collision systems, particularly at high multiplicity, through a plethora of new and precise measurements of different potential observables~\cite{SKPv2,PGhoshv2,dEnterriav2,Wernerv2,Ortizv2,Ortizv22,ALICEpAfilippaper,ATLASJetQuenchingpPb,PHENIX2018lia}. Measurements primarily related to the soft QCD sector of particle production mechanisms have brought to the forefront various observations commonly understood as due to medium formation such as the long-range ridge-like structure at the near side in two-particle angular correlations~\cite{CMSridge,ATLASridge1,ATLASridge2,CMSridge2,CMSridge3}, strangeness enhancement~\cite{ALICEStrangeEnhance1, ALICEStrangeEnhance2,ALICEStrangeEnhance3}, and elliptic flow ($v_{2}$)~\cite{CMSv2,ATLASv2}. In terms of the hard probes the suppression of inclusive jet yield has been measured in different centrality classes of d–Au (deuteron-gold) collisions at RHIC~\cite{PHENIX} and \pPb collisions at the LHC~\cite{ATLAS:2014cpa}. However, no conclusive evidence of jet quenching has been found yet within the current precision achieved in experiments~\cite{ALICERaa1,ALICERaa2,ATLAS:2014cpa}. This leaves the possibility of QGP formation in small collision systems as an open question that must be addressed and investigated further. In view of this, intra-jet properties such as mean charged-particle multiplicity in jets and jet fragmentation functions are promising observables since they are more sensitive to the details of the parton shower and hadronization processes in QCD~\cite{ImpJetShape1,ImpJetShape2,QCDcoherence2,ImpJetProp} compared to inclusive jet observables. In addition to providing a more stringent test of both perturbative and non-perturbative aspects of QCD for minimum-bias (MB) \pp collisions~\cite{CMSjetshape1,ATLASJetshapeAndFrag}, the charged-particle multiplicity and fragmentation functions also serve as potential observables to capture any possible QGP-like effects in high-multiplicity (HM) small collision systems. These QGP-like effects might lead to softening and broadening of jets due to multiple parton scatterings, resulting in modifications of charged-particle multiplicity distributions and fragmentation functions of jets.

Numerous measurements of intra-jet properties have been performed in hadronic collisions. Jet properties were previously measured by the CDF~\cite{CDFjetshapeandUE,CDFjetshape2} and D0~\cite{D0jetshape} Collaborations in p$\overline{\rm p}$ collisions at the Tevatron and recently by the ALICE~\cite{ALICExsec1}, ATLAS~\cite{ATLASjetshape1,ATLASJetshapeAndFrag}, and CMS~\cite{CMSjetshape1,CMSjetshape2} Collaborations in pp collisions at the LHC. Measurements of jet fragmentation functions have also been reported by the CDF Collaboration~\cite{CDFfrag} in p$\overline{\rm p}$ collisions, whereas ALICE~\cite{ALICExsec1,ALICExsec2}, ATLAS~\cite{ATLASJetshapeAndFrag,ATLASjetfrag1,ATLASfrag2}, and CMS~\cite{CMSfrag1} Collaborations have also studied jet fragmentation functions in \pp and \PbPb (lead-lead) collisions at LHC energies. The STAR~\cite{starjetobs} and PHENIX~\cite{PHENIXjetobs,PHENIXjetobsproc} Collaborations have measured jet shape observables and jet fragmentation functions in \AuAu (gold-gold) collisions at RHIC energies.

This article presents the first measurement of the multiplicity dependence of charged-particle intra-jet properties in \pp collisions at \s = 13 TeV. In this study, the average jet constituent multiplicity $\left<N_{\rm ch}\right>$ and the distributions of jet fragmentation function variables, $z^{\rm ch}$ and $\xi^{\rm ch}$ are measured for leading charged-particle jets (jet with the highest \pt in an event) with jet resolution parameters $R$ = 0.2, 0.3, and 0.4 as a function of jet \pt in minimum-bias and high-multiplicity pp collisions using the ALICE detector at the LHC. Leading charged-particle jets are considered since they are theoretically well-defined objects and less prone to experimental effects compared to inclusive jets~\cite{leadjets}. Moreover, the formation and evolution of leading jets can also be described by jet functions which satisfy DGLAP-type evolution equations similar to inclusive jets, and therefore, they are comparable with the QCD hard scattering models~\cite{leadjets}. It is worth mentioning that the selection of leading jets might introduce a surface-bias in jet modification signals in the presence of QGP-like effects in high-multiplicity small collision systems.

This paper is organized as follows: Section~\ref{dataset} describes the experimental setup of the ALICE detector and the data samples used in this study. Details of jet reconstruction and jet observables are discussed in Sec.~\ref{jetreco}. The procedures applied to correct the measured distributions for detector effects and underlying event contaminations are presented in Sec.~\ref{corrections}. Section~\ref{systematics} outlines the estimation of systematic uncertainties from various sources. Results are presented and discussed in detail in comparison with predictions from QCD-inspired event generators in Sec.~\ref{res} and the conclusions are summarized in Sec.~\ref{summary}.
\section{Experimental setup, data sets, and event selection}
\label{dataset}
This analysis uses the data from pp collisions at \s = 13\,TeV collected in 2016, 2017, and 2018 with the ALICE detector at the LHC. The ALICE detector and its performance are described in detail in Refs.~\cite{ALICEExp, ALICEPerf}. Events are selected using the information from two V0 scintillator detectors~\cite{ALICEV0}, V0A and V0C, which cover an azimuthal acceptance of 0 $< \varphi <$ 2$\pi$ and pseudorapidity 2.8 $< \eta <$ 5.1 and --3.7 $< \eta <$ --1.7, respectively.  The online trigger for MB events requires the coincidence of signals both in the V0A and V0C detectors, while the high-multiplicity sample is collected using a HM trigger condition~\cite{FilipPaper} that requires the V0M signal amplitude (sum of V0A and V0C signal amplitudes) to be greater than 5 times of its mean signal amplitude $\langle\rm{V0M}\rangle$ in MB events~\cite{ALICEpartdensity}. 
Charged-particle tracks are reconstructed in the midrapidity region ($|\eta| < $ 0.9) utilizing the information from two central barrel detectors, the Inner Tracking System (ITS) and the Time Projection Chamber (TPC) placed inside a large solenoidal magnet with a uniform magnetic field of $B = 0.5$ T~\cite{ALICEPerf} and field lines along the beam direction. The primary vertex of the collision is reconstructed from tracks. Events with a primary vertex outside $\pm$10 cm along the beam direction from the nominal interaction point are rejected to guarantee a uniform acceptance of the central barrel detectors. Events with collision pileup are removed by rejecting events with multiple reconstructed vertices~\cite{ALICEPerf, ALICEpp13JetXsec}. The results presented in this paper are based on 1832 million MB and 870 million HM events corresponding to integrated luminosities of 32 $\rm{nb}^{-1}$ and 10 $\rm{pb}^{-1}$ respectively.

The analysis is carried out using the primary charged particles, defined as all particles with a mean proper lifetime $\tau >$ 1 cm/$c$, which are either produced directly in the interaction or from decays of particles with mean proper lifetime $\tau <$ 1 cm/$c$~\cite{ALICEPrimary}. Jets are reconstructed from charged-particle tracks measured with the ITS and TPC detectors.
To ensure an approximately uniform azimuthal acceptance and good momentum resolution, charged tracks are reconstructed using a hybrid selection technique~\cite{ALICEpp13JetXsec, ALICEppPbPb5TeVJetSpectra}, where two different classes of tracks are combined. In the first class, tracks are required to include at least one hit in the silicon pixel detector (SPD), which equips the two innermost layers of the ITS. The second class contains tracks without hits in the SPD, where the primary vertex is used as an initial point of the trajectory to improve the estimation of the particle momentum. Tracks with transverse momentum \pt $>$ 0.15 GeV/$c$ in the pseudorapidity range $|\eta| < $ 0.9 over the full azimuth (0 $< \varphi <$ 2$\pi$) are considered in this analysis. The hybrid track reconstruction efficiency in both MB and HM events is found to be about 85\% at \pt = 1 GeV/$c$, decreasing to 74\% at \pt = 50 GeV/$c$. Primary-track momentum resolution is 0.7\% at \pt = 1 GeV/$c$, increasing to 3.7\% at \pt = 50 GeV/$c$.

The corrections for detector effects and the evaluation of systematic uncertainties are performed with the help of simulations based on Monte Carlo (MC) event generators PYTHIA\,8~\cite{Pythia8pt2} with Monash 2013 tune~\cite{Monash} (hereafter referred to as PYTHIA\,8) and EPOS LHC~\cite{EPOSLHC}. PYTHIA\,8 is a standard tool for studying high-energy physics collisions, predominantly based on 2 $\rightarrow$ 2 hard scattering processes. It introduces an impact-parameter-dependent multiparton Interaction (MPI) framework to model the soft underlying event (UE). For the hadronization of partons, it uses the Lund string fragmentation model. In the Monash tune of PYTHIA\,8, parameters relevant to initial-state radiation (ISR) and MPI are tuned using MB, Drell-Yan, and UE data from the Tevatron, SPS and LHC. The EPOS event generator is a parton-based MC model with flux tube initial condition for hadron-hadron collisions. It uses the Gribov-Regge theory to describe soft interactions. The EPOS LHC generator is tuned to LHC data to describe the results from various collision systems at different center-of-mass energies, particularly the observed collective behavior in \pp and \pPb collisions at LHC energies. The selections of MB and HM events in the simulated data are the same as in experimental data.

\section{Jet reconstruction and jet observables}
\label{jetreco}
Charged-particle jets are reconstructed from charged-particle tracks with \pt $>$ 0.15 GeV/$c$ and $|\eta| <$ 0.9 using the anti--$k_{\rm T}$ algorithm~\cite{AntikT} with resolution parameters $R$ = 0.2, 0.3, and 0.4 using FastJet 3.2.1~\cite{FastJet}. The pseudorapidity coverage of the reconstructed jets is limited within the fiducial acceptance of the TPC, $|\eta_{\rm jet}| < (0.9 - R)$, to minimize the TPC edge effects in jet reconstruction. Leading jets within the \pt interval 5--110 GeV/$c$ are considered for this analysis.

The performance of jet reconstruction is studied using PYTHIA\,8. Particles produced directly from the MC event generator (truth-level) are transported through a GEANT~3~\cite{GEANT3} simulation of the ALICE detector system to obtain the reconstructed tracks (detector-level). Truth-level particles (detector-level tracks) are used to reconstruct the truth-level (detector-level) jets by applying the same algorithms and kinematic selections as in data. Detector-level leading jets are matched to the geometrically closest truth-level jets and a one-to-one matching is ensured between them. The axes of the matched jets are required to be within $\Delta R < 0.6\,R$ to minimize unrealistic matching. The matched jets are used to calculate the jet energy scale JES = $(p_{\rm T, det }^{\rm jet,\,ch} - p_{\rm T, truth}^{\rm jet,\,ch}) / p_{\rm T, truth}^{\rm jet,\,ch}$, jet energy resolution JER = $\sigma (p_{\rm T, det }^{\rm jet,\,ch}) / p_{\rm T, truth}^{\rm jet,\,ch}$ (where $\sigma$ is the width of the $p_{\rm T, det }^{\rm jet,\,ch}$ distribution for a given value of $p_{\rm T, truth}^{\rm jet,\,ch}$), and jet reconstruction efficiency $\epsilon_{\rm reco}$ for $R =$ 0.2, 0.3, and 0.4, where $p_{\rm T, truth}^{\rm jet,\,ch}$ and $p_{\rm T, det }^{\rm jet,\,ch}$ denote the transverse momentum of truth- and detector-level jets, respectively. The JES distribution shows a peak at zero with an asymmetric tail towards negative values due to tracking inefficiencies, which is characterized by the mean value of JES denoted as $\Delta_{\rm JES}$.
Table~[\ref{jetperformance}] summarizes the values of $\Delta_{\rm JES}$, JER, and $\epsilon_{\rm reco}$ for different jet-\pt ranges.

\begin{table}[h!]
  \footnotesize
  \begin{center}
    \caption{Approximate values of $\Delta_{\rm JES}$, JER, and $\epsilon_{\rm reco}$ to characterize the jet reconstruction performance for jet $R$ = 0.2, 0.3, and 0.4.}
    \label{jetperformance}
    \renewcommand{\arraystretch}{1.3}
    \begin{tabular}{|c|ccc|ccc|ccc|}
      \hline
      $p_{\rm T, truth}^{\rm jet,\,ch}$ & \multicolumn{3}{c|}{$R = 0.2$} & \multicolumn{3}{c|}{$R = 0.3$}  & \multicolumn{3}{c|}{$R = 0.4$}  \\
      \cline{2-10}
      (GeV/$c$) & $\Delta_{\rm JES} (\%)$ & JER (\%) & $\epsilon_{\rm reco} (\%)$ & $\Delta_{\rm JES} (\%)$ & JER (\%)& $\epsilon_{\rm reco} (\%)$ & $\Delta_{\rm JES} (\%)$ & JER (\%) & $\epsilon_{\rm reco} (\%)$ \\
      \hline
      10--20 &-9 &20& 89&-10 &20 & 90&-12 &20 &91 \\
      20--30 &-11 & 21&94 &-12 & 20&95 &-13&20 & 95\\
      30--40 &-13 &21 &95 &-14 &20 & 96&-14&20 & 96\\
      40--50 &-14 &21&96 &-15& 20&97 &-15& 20&97 \\
      80--90 &-18 &23&97&-19&22 &97 &-18&22 & 97\\
      \hline
    \end{tabular}
  \end{center}
\end{table}

The intra-jet properties such as mean charged-particle multiplicity within a jet cone and the distributions of jet fragmentation function variables, $z^{\rm ch}$ and $\xi^{\rm ch} $ are measured for leading jets in both minimum-bias and high-multiplicity pp collisions. The number of charged particles constituting the jet is termed charged-particle multiplicity $N_{\rm ch}$. The mean charged-particle multiplicity $\left<N_{\rm ch}\right>$ is calculated and presented as a function of the leading jet \pt.
The jet fragmentation function variables, $z^{\rm ch}$ and $\xi^{\rm ch} $ are defined as:
\begin{equation}
   z^{\rm ch} = \frac{p_{\rm T}^{\rm particle}}{p_{\rm T}^{\rm jet,\,ch}}, 
\end{equation}
\begin{equation}
   \xi^{\rm ch} = {\rm ln} \left(\frac{1}{z^{\rm ch}}\right), 	
\end{equation}
where $p_{\rm T}^{\rm particle}$ is the transverse momentum of the jet constituent. The distributions are normalized by the total number of leading jets and explicitly describe the energy sharing between constituents within a jet. The $\xi^{\rm ch} $ distribution is complementary to $z^{\rm ch}$, which emphasizes fragmentation into low momentum constituents and is particularly suited to illustrate the QCD coherence effects~\cite{QCDcoherence3, QCDcoherence4, QCDcoherence1, QCDcoherence2, QCDcoherence5, QCDcoherence6}.

\section{Corrections}
\label{corrections}
\subsection{Unfolding}
\label{unf}
The measured distributions are corrected for detector effects such as limited track reconstruction efficiency, finite track-\pt resolution, and particle--material interactions using an iterative method based on Bayes' theorem~\cite{BayesianUnf} implemented in the RooUnfold package~\cite{RooUnfold}. To account for these effects, a 4D response matrix (RM) is constructed from simulated data and considered as an input to the unfolding procedure, which maps between the truth- and detector-level jet observables. Before the construction of the response matrix, the jets at the truth- and detector-level are matched as described in Sec.~\ref{jetreco}. The elements of the 4D response matrix are $p_{\rm T,\,det}^{\rm jet,\,ch}$, $Obs_{\rm det}$, $p_{\rm T,\,truth}^{\rm jet,\,ch}$ and $Obs_{\rm truth}$,  where $p_{\rm T, \,det}^{\rm jet,\,ch}$ and $p_{\rm T, \,truth}^{\rm jet,\,ch}$ are detector- and truth-level jet \pt and $Obs_{\rm det}$ and $Obs_{\rm truth}$ stand for the observables, $Obs\in \{N_{\rm ch}, z^{\rm ch}, \xi^{\rm ch}\}$. For $z^{\rm ch}$ and $\xi^{\rm ch}$, the truth- and detector-level jet constituents are also matched before constructing the response matrices. Any detector-level (truth-level) jet constituent without an associated matched truth-level (detector-level) jet constituent is termed fake (miss) and is fed to the response matrix in addition to the matched jet constituents to account for the efficiency and purity of the constituent matching procedure. The unfolded distributions obtained using the Bayesian unfolding technique primarily depend on two important factors, the regularization parameter and the prior distribution. In the case of Bayesian unfolding, the regularization parameter is the number of iterations. The regularization parameter is tuned to reduce the variance of the unfolded distribution. The truth-level distributions are provided as the prior in the unfolding process that gets updated in subsequent iterations.

Two types of closure tests are performed to validate the unfolding procedure, known as statistical and shape closure tests. In the statistical closure test, two statistically independent simulated datasets are considered, where the response matrix is built from one sample, and the truth- and detector-level distributions of $N_{\rm ch}$, $z^{\rm ch}$, and $\xi^{\rm ch}$ are obtained from the other sample. The detector-level distributions are then unfolded and compared with the truth-level distribution to check the robustness of the unfolding procedure against the statistical fluctuations in the data. In the shape closure test, a similar approach as the statistical closure is applied, however, the response matrix is reweighted with the ratio between the measured distribution and the one from detector-level MC. Then, the unfolded distribution is compared with truth-level distributions to check the robustness of the unfolding against the change in the shape of distributions. Proper closure is found in both tests within the statistical uncertainties.

\subsection{Underlying event subtraction}
\label{ue_sub}
The underlying event (UE) consists of all particles produced in the collision that are not an integral part of the jet or produced directly from the hard scattering. In pp collisions, some of the important sources of UE are beam remnants, multiparton interactions, initial- and final-state radiations. The perpendicular cone method used in Refs.~\cite{ALICExsec1,ALICExsec2} is adopted to estimate UE and correct the corresponding distributions of jet observables in both MB and HM events. 

In this approach, the UE particle yield is measured event-by-event within a cone of the same radius as the jet resolution parameter located at the same pseudorapidity as the leading jet, but in the direction perpendicular to the leading jet axis. The information of particles within the perpendicular cone is used to estimate the UE contributions to the jet observables.

The UE distributions of $N_{\rm ch}$, $z^{\rm ch}$, and $\xi^{\rm ch}$ are corrected for the detector effects using the unfolding procedure discussed in Sec.~\ref{unf}. After unfolding, the unfolded UE distributions are subtracted from the unfolded signal distributions on a statistical basis, however, a simultaneous correction for the UE contribution to the jet transverse momentum is not applied here~\cite{ALICExsec1,ALICExsec2}. The estimated UE contribution for $\langle N_{\rm ch}\rangle$ in MB events is comparable with the values reported in Ref.~\cite{ALICEUEpp13}.

\section{Systematic uncertainties}
\label{systematics}
The systematic uncertainties associated with the unfolded distributions are mainly arising from the uncertainties in track reconstruction efficiency, the unfolding procedure (variation in the regularization parameter of unfolding, change of prior distribution, and bin truncation), the choice of MC model in the correction procedure, and the uncertainty in the estimation of the UE. For each of these sources, a modified response matrix that incorporates the variation due to the respective uncertainties is built (as described below) and used to unfold the measured distribution. The difference between the corrected distributions unfolded with the default and modified response matrices is quoted as the corresponding systematic uncertainty. The total systematic uncertainty is calculated by taking the quadrature sum of all the individual sources, assuming that all the sources are uncorrelated.

\begin{table}[h!]
	\scriptsize
	\caption{Summary of systematic uncertainties (in \%) on $\langle N_{\rm ch} \rangle$ for selected intervals of jet \pt for jet $R$ = 0.2, 0.3, and 0.4 in MB and HM events.}
	\label{SysTable_Nch}
	\begin{center}
		\begin{threeparttable}
			\renewcommand{\arraystretch}{1.6}
			\begin{tabular}{||c|c c c|c c c| c c c||}		
				\hline
				\multirow{4}{*}{\textbf{Sources}}  & \multicolumn{9}{c||}{\textbf{Systematic uncertainties on $\langle N_{\rm ch} \rangle$ for MB (\%)}}  \\
				\cline{2-10}
				& \multicolumn{3}{c|}{\textbf{$R = 0.2$}}  & \multicolumn{3}{c|}{\textbf{$R = 0.3$}} & \multicolumn{3}{c||}{\textbf{$R = 0.4$}}\\
				\cline{2-10}
				& \multicolumn{3}{c|}{Jet \pt in GeV/$c$} & \multicolumn{3}{c|}{Jet \pt in GeV/$c$} & \multicolumn{3}{c||}{Jet \pt in GeV/$c$}\\
				\cline{2-10}
				& 5--10 & 45--50 & 90--110 & 5--10 & 45--50 & 90--110 & 5--10 & 45--50 & 90--110 \\
				\cline{1-10}
				\textbf{Track reconst. efficiency} & 0.9 & 2.0 & 2.5 & 1.3 & 2.1 & 2.2 & 1.8 & 2.5 & 2.4\\
				\textbf{Unfolding parameter} & negl. & 0.1 & negl. & 0.1 & 0.1 & 0.2 & 0.1 & negl. & 0.1\\
				\textbf{Prior change} & negl. & 0.5 & negl. & 0.1 & 0.2 & 0.3 & 0.1 & 0.2 & 0.2\\							
				\textbf{Bin truncation} & 10.4 & 0.3 & 2.2 & 11.3 & 0.3 & 1.2 & 11.1 & 0.3 & 1.1\\
				\textbf{MC generator} & 1.0 & 1.4 & 10.2 & 1.0 & 1.9 & 3.2 & 0.8 & 2.2 & 4.0\\	
				\textbf{UE} & 0.1 & 0.1 & negl. & 0.3 & 0.2 & 0.7 & 0.7 & 0.4 & 0.8\\
				\hline
				\textbf{Total} & 10.5 & 2.5 & 10.7 & 11.4 & 2.9 & 4.2 & 11.3 & 3.4 & 4.9\\
				\hline				
				\hline
				\multirow{4}{*}{\textbf{Sources}}  & \multicolumn{9}{c||}{\textbf{Systematic uncertainties on $\langle N_{\rm ch} \rangle$ for HM (\%)}}  \\
                          \cline{2-10}
                                                                    & \multicolumn{3}{c|}{\textbf{$R = 0.2$}}  & \multicolumn{3}{c|}{\textbf{$R = 0.3$}} & \multicolumn{3}{c||}{\textbf{$R = 0.4$}}\\
                          \cline{2-10}
                                                                    & \multicolumn{3}{c|}{Jet \pt in GeV/$c$} & \multicolumn{3}{c|}{Jet \pt in GeV/$c$} & \multicolumn{3}{c||}{Jet \pt in GeV/$c$}\\
                          \cline{2-10}
                                                                    & 5--10 & 45--50 & 90--110 & 5--10 & 45--50 & 90--110 & 5--10 & 45--50 & 90--110 \\
                          \cline{1-10}
                          \textbf{Track reconst. efficiency} & 1.1 & 1.5 & 2.4 & 2 & 2.3 & 2.6 & 2.5 & 2.4 & 3.4\\
                          \textbf{Unfolding parameter} & negl. & 0.1 & 0.1 & negl. & 0.1 & 0.1 & 0.1 & 0.1 & 0.1\\
                          \textbf{Prior change} & negl. & 0.1 & 0.3 & negl. & 0.1 & 1.4 & 0.2 & 0.1 & 0.4\\
                          \textbf{Bin truncation} & 3.6 & 0.3 & 0.7 & 8.7 & 0.2 & 1.3 & 4.1 & 0.3 & 0.8\\
                          \textbf{MC generator} & 1.0 & 1.4 & 10.2 & 1.0 & 1.9 & 3.2 & 0.8 & 2.2 & 4.0\\	
                          \textbf{UE} & 0.7 & 0.4 & negl. & 2.1 & 0.8 & 0.3 & 3.2 & 1.3 & 1.0\\
				\hline
                          \textbf{Total} & 4.0 & 2.1 & 10.5 & 9.2 & 3.1 & 4.5 & 5.8 & 3.5 & 5.4\\
				\hline
			\end{tabular}
		\end{threeparttable}	
	\end{center}	
\end{table}

\begin{table}[h!]
	\scriptsize
	\caption{Summary of systematic uncertainties (in \%) on ${\rm d}N/{\rm d}z^{\rm ch}$ in $z^{\rm ch}$ bins for selected intervals of jet \pt for jet $R$ = 0.2, 0.3, and 0.4 in MB and HM events.}
	\label{SysTable_zCh}
	\begin{center}
		\begin{threeparttable}
			\renewcommand{\arraystretch}{1.6}
			\begin{tabular}{||c|c|c c c|c c c| c c c||}		
				\hline
				\multirow{3}{*}{\textbf{Jet \pt}} & \multirow{4}{*}{\textbf{Sources}}  & \multicolumn{9}{c||}{\textbf{Systematic uncertainties on ${\rm d}N/{\rm d}z^{\rm ch}$ for MB (\%)}}  \\
				\cline{3-11}
                                                                    &  & \multicolumn{3}{c|}{\textbf{$R = 0.2$}}  & \multicolumn{3}{c|}{\textbf{$R = 0.3$}} & \multicolumn{3}{c||}{\textbf{$R = 0.4$}}\\
                          \cline{3-11}
                          (GeV/$c$)& & \multicolumn{3}{c|}{$z^{\rm ch}$ bin} & \multicolumn{3}{c|}{$z^{\rm ch}$ bin} & \multicolumn{3}{c||}{$z^{\rm ch}$ bin}\\
                          \cline{3-11}
                                                                    & & 0 -- 0.1 & 0.3 -- 0.4 & 0.9 -- 1 & 0 -- 0.1 & 0.3 -- 0.4 & 0.9 -- 1 & 0 -- 0.1 & 0.3 -- 0.4 & 0.9 -- 1 \\
                          \cline{1-11}
                          \multirow{6}{*}{10--20}&\textbf{Track reconst. efficiency} & 4.4 & 1.2 & 4.7 & 4.6 & 0.4 & 6.6 & 4.5 & 0.3 & 8.4\\
                                                                    &\textbf{Unfolding parameter} & 0.8 & 0.1 & 0.1 & 0.7 & 0.1 & 0.2 & 0.8 & 0.3 & 0.1\\				
                                                                    &\textbf{Prior change} & 3.8 & 0.6 & 4.7 & 2.3 & 2.1 & 3.1 & 2.1 & 2.1 & 2.3\\		
                                                                    &\textbf{Bin truncation} & 5.4 & 5.8 & 16.4 & 8.9 & 7.9 & 22.2 & 12.2 & 10.3 & 27.9\\		
                                                                    &\textbf{MC generator} & 5.1 & 1.1 & 11.4 & 2.2 & 0.9 & 9.4 & 0.6 & 0.6 & 8.0\\
                                                                    &\textbf{UE} & 4.5 & 0.1 & negl. & 3.7 & 0.2 & negl. & 2.9 & 0.2 & 0.1\\
                          \hline
                                                                    &\textbf{Total} & 10.5 & 6.1 & 21.1 & 11.2 & 8.2 & 25.2 & 13.5 & 10.5 & 30.3\\
                          \hline
                          \hline
                          \multirow{6}{*}{60--80}&\textbf{Track reconst. efficiency} & 3.2 & 0.8 & 12.6 & 3.2 & 1.5 & 14.4 & 3.5 & 1.7 & 16.3\\
                                                                    &\textbf{Unfolding parameter} & 0.3 & 1.1 & 1.8 & 0.4 & 1.7 & 3.4 & 0.4 & 0.6 & 0.6\\
                                                                    &\textbf{Prior change} & 1.4 & 0.9 & 6.5 & 1.8 & 2.1 & 17.5 & 0.4 & 1.3 & 3.4\\
                                                                    &\textbf{Bin truncation} & 0.9 & 0.5 & 0.5 & 0.6 & 0.4 & negl. & 0.2 & 0.5 & 0.4\\
                                                                    &\textbf{MC generator} & 0.4 & 15.4 & 35.9 & 10.8 & 23.6 & 29.4 & 16.4 & 17.5 & 60.3\\
                                                                    &\textbf{UE} & 1.7 & negl. & negl. & 2.9 & 0.1 & negl. & 3.3 & 0.1 & negl.\\
                          \hline
                                                                    &\textbf{Total} & 4.0 & 15.5 & 38.6 & 11.8 & 23.8 & 37.3 & 17.1 & 17.6 & 62.6\\
                          \hline								
                          \hline
                          \multirow{3}{*}{\textbf{Jet \pt}} & \multirow{4}{*}{\textbf{Sources}}  & \multicolumn{9}{c||}{\textbf{Systematic uncertainties on ${\rm d}N/{\rm d}z^{\rm ch}$ for HM (\%)}}  \\
                          \cline{3-11}
                                                                    &  & \multicolumn{3}{c|}{\textbf{$R = 0.2$}}  & \multicolumn{3}{c|}{\textbf{$R = 0.3$}} & \multicolumn{3}{c||}{\textbf{$R = 0.4$}}\\
                          \cline{3-11}
                          (GeV/$c$)& & \multicolumn{3}{c|}{$z^{\rm ch}$ bin} & \multicolumn{3}{c|}{$z^{\rm ch}$ bin} & \multicolumn{3}{c||}{$z^{\rm ch}$ bin}\\
                          \cline{3-11}
                                                                    & & 0 -- 0.1 & 0.3 -- 0.4 & 0.9 -- 1 & 0 -- 0.1 & 0.3 -- 0.4 & 0.9 -- 1 & 0 -- 0.1 & 0.3 -- 0.4 & 0.9 -- 1 \\
                          \cline{1-11}
                          \multirow{6}{*}{10--20}&\textbf{Track reconst. efficiency} & 5.8 & 0.9 & 7.3 & 6.1 & 0.9 & 8.8 & 4.4 & 1.3 & 15.7\\
                                                                    &\textbf{Unfolding parameter} & 0.1 & negl. & 0.4 & 0.2 & 0.1 & 0.2 & 0.8 & 0.2 & 0.1\\
                                                                    &\textbf{Prior change} & 0.8 & 0.9 & 4.6 & negl. & 1.1 & 2.6 & 1.8 & 2.0 & 0.8\\
                                                                    &\textbf{Bin truncation} & 10.9 & 7.4 & 19.0 & 15.1 & 10.3 & 25.7 & 19.5 & 14.0 & 28.7\\
                                                                    &\textbf{MC generator} & 5.1 & 1.1 & 11.4 & 2.2 & 0.9 & 9.4 & 0.6 & 0.6 & 8.0\\
                                                                    &\textbf{UE} & 0.6 & 0.1 & 4.2 & 0.1 & 0.1 & 0.6 & 0.5 & 0.1 & 1.6\\
                          \hline
                                                                    &\textbf{Total} & 13.4 & 7.6 & 24.1 & 16.4 & 10.4 & 28.9 & 20.1 & 14.2 & 33.7\\
                          \hline
                          \hline
                          \multirow{6}{*}{60--80}&\textbf{Track reconst. efficiency} & 3.5 & 1.3 & 4.6 & 3.9 & 0.8 & 3.2 & 4.2 & 1.3 & 16.0\\
                                                                    &\textbf{Unfolding parameter} & 0.3 & 0.1 & 4.4 & 0.2 & 0.4 & 1.3 & 0.2 & 0.3 & 2.5\\
                                                                    &\textbf{Prior change} & 2.3 & 0.5 & 5.4 & 2.3 & 0.8 & 8.8 & 2.5 & 0.5 & 14.0\\
                                                                    &\textbf{Bin truncation} & 1.5 & 1.2 & 0.7 & 1.3 & 0.8 & 0.1 & 0.9 & 0.6 & 0.2\\
                                                                    &\textbf{MC generator} & 0.4 & 15.4 & 35.9 & 10.8 & 23.6 & 29.4 & 16.4 & 17.5 & 60.3\\
                                                                    &\textbf{UE} & 0.9 & negl. & negl. & 1.5 & 0.1 & negl. & 1.7 & 0.1 & negl.\\
                          \hline
                                                                    &\textbf{Total} & 4.6 & 15.5 & 36.9 & 11.9 & 23.6 & 30.9 & 17.2 & 17.6 & 64.0\\
                          \hline				
			\end{tabular}
                      \end{threeparttable}	
                    \end{center}	
                  \end{table}

                  \begin{table}[h!]
                    \scriptsize
                    \caption{Summary of systematic uncertainties (in \%) on ${\rm d}N/{\rm d}\xi^{\rm ch}$ in $\xi^{\rm ch}$ bins for selected intervals of jet \pt for jet $R$ = 0.2, 0.3, and 0.4 in MB and HM events.}
                    \label{SysTable_xiCh}
                    \begin{center}
                      \begin{threeparttable}
			\renewcommand{\arraystretch}{1.6}
			\begin{tabular}{||c|c|c c c|c c c| c c c||}		
                          \hline
                          \multirow{3}{*}{\textbf{Jet \pt}} & \multirow{4}{*}{\textbf{Sources}}  & \multicolumn{9}{c||}{\textbf{Systematic uncertainties on ${\rm d}N/{\rm d}\xi^{\rm ch}$ for MB (\%)}}  \\
                          \cline{3-11}
                                                              &  & \multicolumn{3}{c|}{\textbf{$R = 0.2$}}  & \multicolumn{3}{c|}{\textbf{$R = 0.3$}} & \multicolumn{3}{c||}{\textbf{$R = 0.4$}}\\
                          \cline{3-11}
                          (GeV/$c$)& & \multicolumn{3}{c|}{$\xi^{\rm ch}$ bin} & \multicolumn{3}{c|}{$\xi^{\rm ch}$ bin} & \multicolumn{3}{c||}{$\xi^{\rm ch}$ bin}\\
                          \cline{3-11}
                                                              & & 0 -- 0.4 & 2.8 -- 3.2 & 4.8 -- 5.2 & 0 -- 0.4 & 2.8 -- 3.2 & 4.8 -- 5.2 & 0 -- 0.4 & 2.8 -- 3.2 & 4.8 -- 5.2 \\
                          \cline{1-11}
                          \multirow{6}{*}{10--20}&\textbf{Track reconst. efficiency} & 3.1 & 4.3 & 5.7 & 4.4 & 4.4 & 20.6 & 6.0 & 4.3 & 18.0\\
				&\textbf{Unfolding parameter} & 0.3 & 0.6 & 21.5 & 0.2 & 0.6 & 15.0 & 0.1 & 0.8 & 16.6\\
                                                              &\textbf{Prior change} & 1.0 & 3.2 & 1.5 & 0.9 & 2.2 & 9.8 & negl. & 0.6 & 6.9\\
                                                              &\textbf{Bin truncation} & 12.7 & 5.4 & 1.6 & 17.2 & 8.9 & 13.7 & 21.7 & 12.0 & 28.8\\
                                                              &\textbf{MC generator} & 5.3 & 4.7 & 28.2 & 7.2 & 1.7 & 13.6 & 9.2 & 0.7 & 9.0\\
                                                              &\textbf{UE} & negl. & 1.5 & 5.6 & 0.1 & 1.4 & 3.8 & 0.1 & 0.8 & 2.8\\
                          \hline
                                                              &\textbf{Total} & 14.1 & 9.1 & 36.4 & 19.2 & 10.4 & 33.7 & 24.3 & 12.8 & 39.6\\
                          \hline
                          \hline
                          \multirow{6}{*}{60--80}&\textbf{Track reconst. efficiency} & 7.4 & 2.9 & 6.9 & 8.8 & 2.8 & 6.2 & 9.1 & 2.7 & 6.1\\
                                                              &\textbf{Unfolding parameter} & 0.4 & 0.3 & 1.9 & 2.3 & 0.3 & 0.4 & 0.4 & 0.6 & 1.4\\
                                                              &\textbf{Prior change} & 1.9 & 0.1 & 13.3 & 5.2 & 3.6 & 15.5 & 6.5 & 5.6 & 12.6\\
                                                              &\textbf{Bin truncation} & 0.2 & 0.3 & 4.4 & negl. & 0.1 & 2.8 & 0.1 & 0.1 & 1.3\\
                                                              &\textbf{MC generator} & 20.4 & 1.8 & 28.3 & 18.0 & 6.8 & 3.2 & 15.5 & 13.4 & 6.0\\
                                                              &\textbf{UE} & negl. & 0.5 & 0.7 & negl. & 0.7 & 1.1 & negl. & 0.8 & 1.6\\
                          \hline
                                                              &\textbf{Total} & 21.8 & 3.5 & 32.4 & 20.8 & 8.2 & 17.3 & 19.1 & 14.8 & 15.4\\
                          \hline								
                          \hline
                          \multirow{3}{*}{\textbf{Jet \pt}} & \multirow{4}{*}{\textbf{Sources}}  & \multicolumn{9}{c||}{\textbf{Systematic uncertainties on ${\rm d}N/{\rm d}\xi^{\rm ch}$ for HM (\%)}}  \\
                          \cline{3-11}
                                                              &  & \multicolumn{3}{c|}{\textbf{$R = 0.2$}}  & \multicolumn{3}{c|}{\textbf{$R = 0.3$}} & \multicolumn{3}{c||}{\textbf{$R = 0.4$}}\\
                          \cline{3-11}
                          (GeV/$c$)& & \multicolumn{3}{c|}{$\xi^{\rm ch}$ bin} & \multicolumn{3}{c|}{$\xi^{\rm ch}$ bin} & \multicolumn{3}{c||}{$\xi^{\rm ch}$ bin}\\
                          \cline{3-11}
                                                              & & 0 -- 0.4 & 2.8 -- 3.2 & 4.8 -- 5.2 & 0 -- 0.4 & 2.8 -- 3.2 & 4.8 -- 5.2 & 0 -- 0.4 & 2.8 -- 3.2 & 4.8 -- 5.2 \\
                          \cline{1-11}
                          \multirow{6}{*}{10--20}&\textbf{Track reconst. efficiency} & 4.2 & 8.7 & 16.3 & 5.0 & 5.2 & 5.8 & 5.0 & 3.8 & 19.9\\
                                                              &\textbf{Unfolding parameter} & 0.6 & 0.2 & 22.2 & 1.1 & 0.1 & 1.5 & 1.1 & 0.7 & 11.3\\
                                                              &\textbf{Prior change} & 0.5 & 0.6 & 7.4 & 1.1 & 0.2 & 4.7 & 2.3 & 2.3 & 3.4\\
                                                              &\textbf{Bin truncation} & 15.1 & 10.2 & 25.2 & 22.1 & 14.7 & 33.6 & 28.0 & 18.7 & 36.6\\
                                                              &\textbf{MC generator} & 5.3 & 4.7 & 28.2 & 7.2 & 1.7 & 13.6 & 9.2 & 0.7 & 9.0\\
                                                              &\textbf{UE} & negl. & 2.0 & 5.0 & 0.1 & 1.8 & 3.5 & 0.1 & 1.1 & 0.1\\
                          \hline
                                                              &\textbf{Total} & 16.6 & 14.4 & 47.6 & 23.8 & 15.8 & 37.2 & 30.0 & 19.3 & 44.2\\
                          \hline
                          \hline
                          \multirow{6}{*}{60--80}&\textbf{Track reconst. efficiency} & 5.9 & 2.9 & 5.4 & 7.3 & 3.3 & 6.4 & 8.0 & 3.1 & 6.3\\
                                                              &\textbf{Unfolding parameter} & 0.8 & 0.2 & 2.1 & 0.6 & 0.1 & 2.2 & 0.6 & negl. & 2.1\\
                                                              &\textbf{Prior change} & 0.9 & 2.4 & negl. & 1.3 & 2.3 & 0.2 & 2.5 & 2.5 & 0.1\\
                                                              &\textbf{Bin truncation} & 0.3 & 1.2 & 3.3 & 0.2 & 1.0 & 2.7 & 0.2 & 0.7 & 1.8\\
                                                              &\textbf{MC generator} & 20.4 & 1.8 & 28.3 & 18.0 & 6.8 & 3.2 & 15.5 & 13.4 & 6.0\\
                                                              &\textbf{UE} & negl. & 0.1 & 0.3 & negl. & negl. & 0.4 & negl. & 0.1 & negl.\\
                          \hline
                                                              &\textbf{Total} & 21.3 & 4.3 & 29.1 & 19.5 & 8.0 & 8.0 & 17.6 & 14.0 & 9.1\\
                          \hline				
			\end{tabular}
                      \end{threeparttable}	
                    \end{center}	
                  \end{table}

The uncertainty on the track reconstruction efficiency is estimated to be 3\% based on variations of track selection criteria and possible imperfections in the description of the TPC--ITS track matching efficiency in the simulation~\cite{ALICExsec4}. Consequently, a new response matrix is constructed after removing 3\% of detector-level tracks randomly before jet finding and is used to unfold the measured data in order to estimate the systematic uncertainties on the reported jet observables.

The number of iterations is optimized to a value that minimizes the total uncertainty in unfolded data. As a systematic study, the number of iterations is varied by $\pm$ 1 with respect to the default value and the average difference of the modified unfolded distributions from the default one is considered a systematic uncertainty. To estimate the systematic uncertainty due to the change in the shape of the prior distribution, the response matrix is reweighted with the ratio between the measured distribution and the one from detector-level MC. The systematic uncertainty is evaluated as the difference between the distributions obtained by unfolding with the default and reweighted response matrices. Additionally, the sensitivity of the unfolded result to the boundary values of the jet \pt interval considered in the response matrix is reflected in bin migration effects and the corresponding systematic uncertainty is estimated by varying the lower bound of detector-level jet \pt by +5 GeV/$c$ before the construction of the modified response matrix as followed in Refs.~\cite{ALICE2021njq,ALICE2022rdg,ALICE2022hyz}. The upper bound of detector-level jet \pt is also simultaneously varied by -20 GeV/$c$ to check the effect on the corrected distributions for larger bin migration.

As discussed in Sec.~\ref{unf}, the response matrices used to unfold the data are constructed using the information of correspondence between truth- and detector-level jets and their constituents obtained from simulations with the PYTHIA\,8 generator. However, the particular structure of jets simulated by one event generator may be different from that in other event generators, which may, in turn, affect the unfolded distributions. To account for the model dependence uncertainty, another MC event generator, EPOS LHC~\cite{EPOSLHC}, is used to construct a modified response matrix, and the corresponding systematic uncertainty is evaluated from the difference with respect to the default results obtained with PYTHIA\,8.

To estimate the systematic uncertainty due to the UE estimation method, the random cone method is applied where two cones are randomly generated with the same pseudorapidity as the leading jet, and with azimuthal angles with respect to the leading jet axis ($\Delta\varphi$) within $\pi/3 < \Delta\varphi < 2\pi/3$ and $-2\pi/3 < \Delta\varphi < -\pi/3$, instead of using a fixed azimuthal angle of $\Delta\varphi = \pi/2$ as done in the perpendicular cone method.
Similarly to the approach adopted in the perpendicular cone method, the UE contributions to the jet observables are estimated using the information of particles from the two random cones and are provided as input to construct the modified response matrices. The difference between the unfolded distributions obtained for the two UE estimation methods is reported as the corresponding systematic uncertainty.

Table~\ref{SysTable_Nch} summarizes the estimated systematic uncertainties on $\langle N_{\rm ch}\rangle$ from the different sources in MB and HM events. Similarly, the systematic uncertainties on $z^{\rm ch}$ and $\xi^{\rm ch}$ distributions in MB and HM events are outlined in Tables~\ref{SysTable_zCh} and~\ref{SysTable_xiCh}, respectively. In most of the cases, the uncertainties due to track reconstruction efficiency and model dependence turn out to be the dominant sources of systematic uncertainties.

\section{Results}
\label{res}
\subsection{Mean charged-particle multiplicity in the leading jet $\left<N_{\rm ch}\right> $}
\begin{figure}[h!]
  \centering
  \includegraphics[scale=0.8]{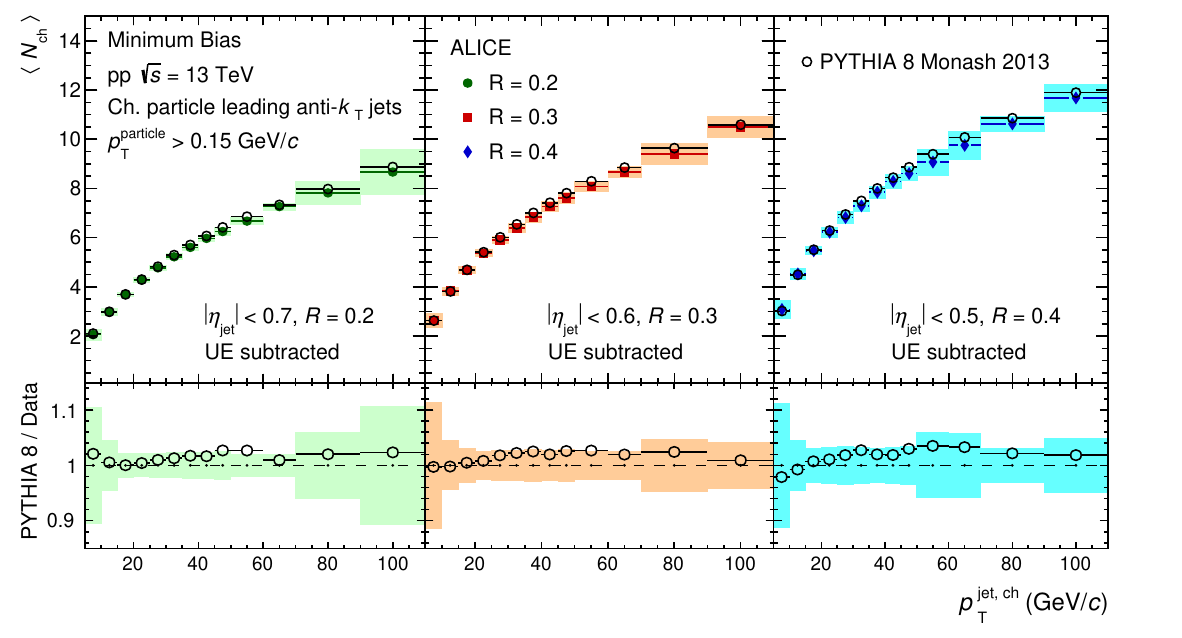}
  \includegraphics[scale=0.8]{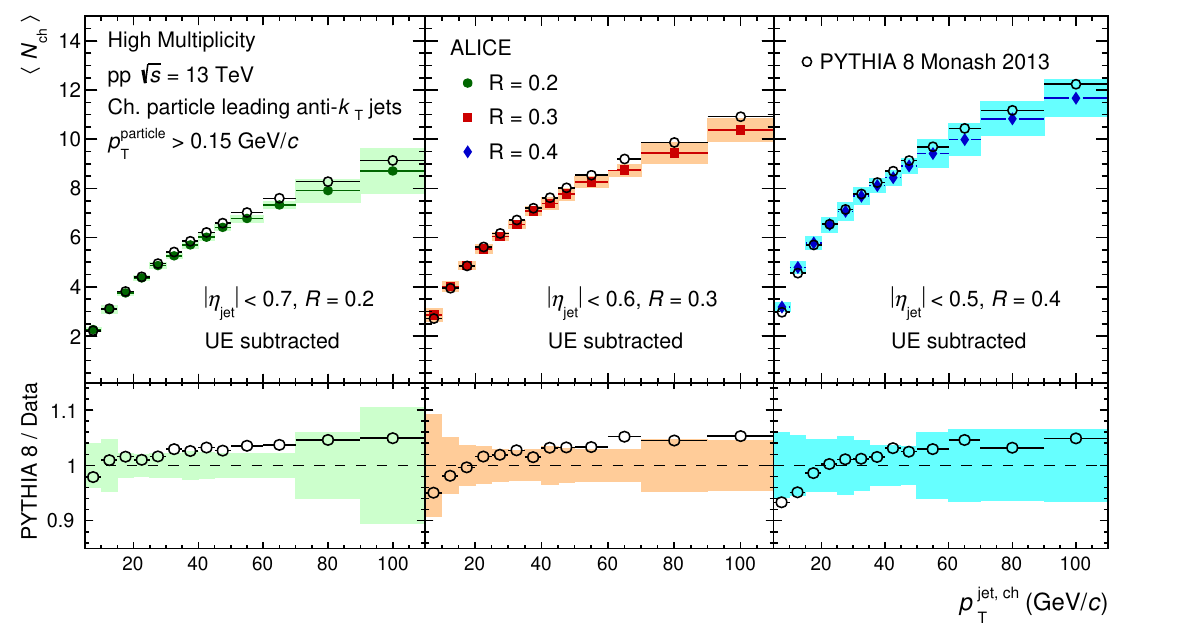}
  \caption{$\left<N_{\rm ch}\right>$ as a function of leading jet \pt for MB (top) and HM (bottom) events for jet radii $R$
    = 0.2 (left), 0.3 (middle), and 0.4 (right). The distributions are
    compared with PYTHIA\,8 predictions.}
  \label{NchUESubFinal_AllR_MBandHM}
\end{figure}

Figure~\ref{NchUESubFinal_AllR_MBandHM} shows the mean number of charged particles within leading jets as a function of jet \pt in pp collisions at \s = 13 TeV for MB (top) and HM (bottom) events. The upper panels show the corrected $\left<N_{\rm ch}\right>$ distributions for jet $R$ = 0.2 (left), 0.3 (middle), and 0.4 (right) in the pseudorapidity ranges $|\eta_{\rm jet}| < (0.9 - R)$. The data points and the corresponding systematic uncertainties are presented by solid markers and shaded bands, respectively. The statistical uncertainties are represented by vertical error bars (smaller than the marker size). Results are compared to predictions from PYTHIA\,8 denoted by open markers. The lower panels show the ratio between PYTHIA\,8 predictions and data. A monotonic increase of $\left<N_{\rm ch}\right>$ is observed with increasing jet \pt as well as with jet radius $R$ for both MB and HM events. The slope of increase at low jet \pt is larger than that at high jet \pt indicating that as \pt increases, more momentum is carried by single constituents. Within systematic uncertainties, PYTHIA\,8 is consistent with the measured trend.

\begin{figure}[h!]
  \centering
  \includegraphics[scale=0.8]{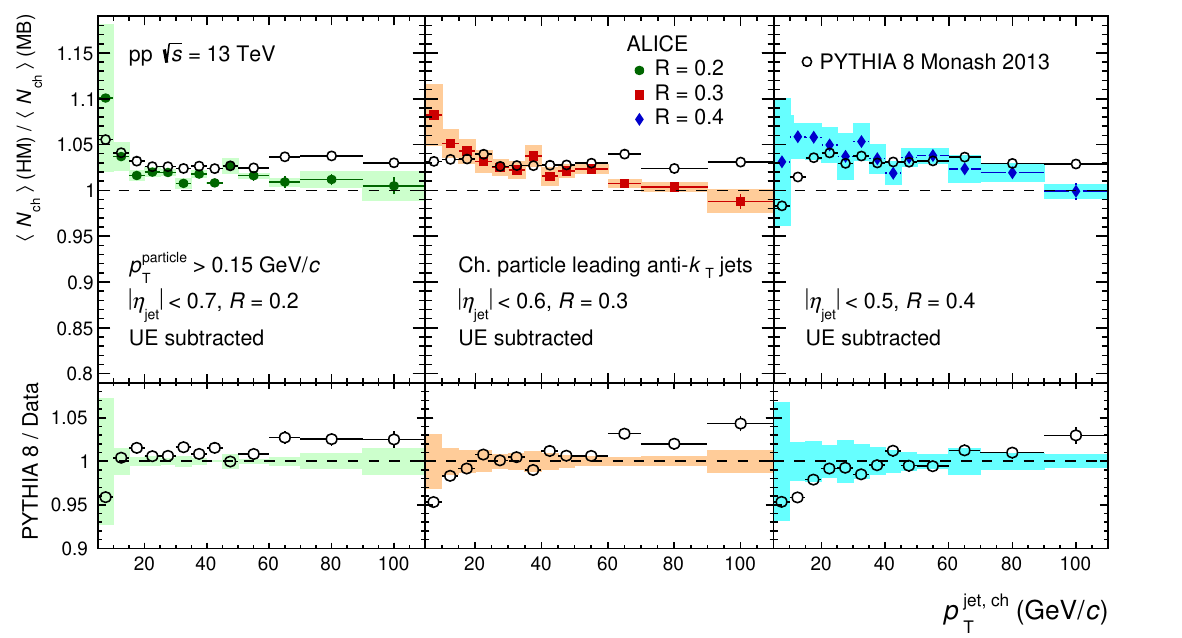}
  \caption{Top panel: The ratio of $\left<N_{\rm ch}\right>$ between
    HM and MB events for jet radii $R$ = 0.2 (left), 0.3 (middle), and 0.4 (right) compared to PYTHIA\,8 predictions. Bottom panel: Ratio between PYTHIA\,8 predictions and the measured values.}
  \label{NchUESubFinal_AllR_HMbyMB}
\end{figure}

The top panels of Fig.~\ref{NchUESubFinal_AllR_HMbyMB} show the ratios of $\left<N_{\rm ch}\right>$ between HM and MB events as a function of jet \pt in comparison to predictions from PYTHIA\,8. The data points are shown by solid markers and the PYTHIA\,8 predictions are represented by open markers for jet $R$ = 0.2 (left), 0.3 (middle), and 0.4 (right). The ratios between PYTHIA\,8 predictions and data are shown in the bottom panels. A mild enhancement in the mean number of charged jet constituents is observed in HM compared to that in MB event class. The magnitude of the enhancement is found to decrease gradually with increasing jet \pt. A maximum increase of $\sim 10\% (8\%, 6\%)$ for jet $R$ = 0.2 (0.3, 0.4) is observed towards low jet \pt while there is no increase at high jet \pt for all $R$. PYTHIA\,8 qualitatively reproduces the data, however, fails to quantitatively reproduce the jet-\pt dependence.
This observation indicates a softening of charged jet constituents in HM events compared to MB for low-\pt jets, which aligns with the CMS measurement of a complementary observable, namely the mean $p_{\rm T}$ of charged jet constituents, in pp collisions at \s = 7 TeV~\cite{CMSmeanpt}.


\subsection{Jet fragmentation}
\subsubsection{$z^{\rm ch}$ distributions}
\begin{figure}[h!]
  \centering
  \includegraphics[scale=0.8]{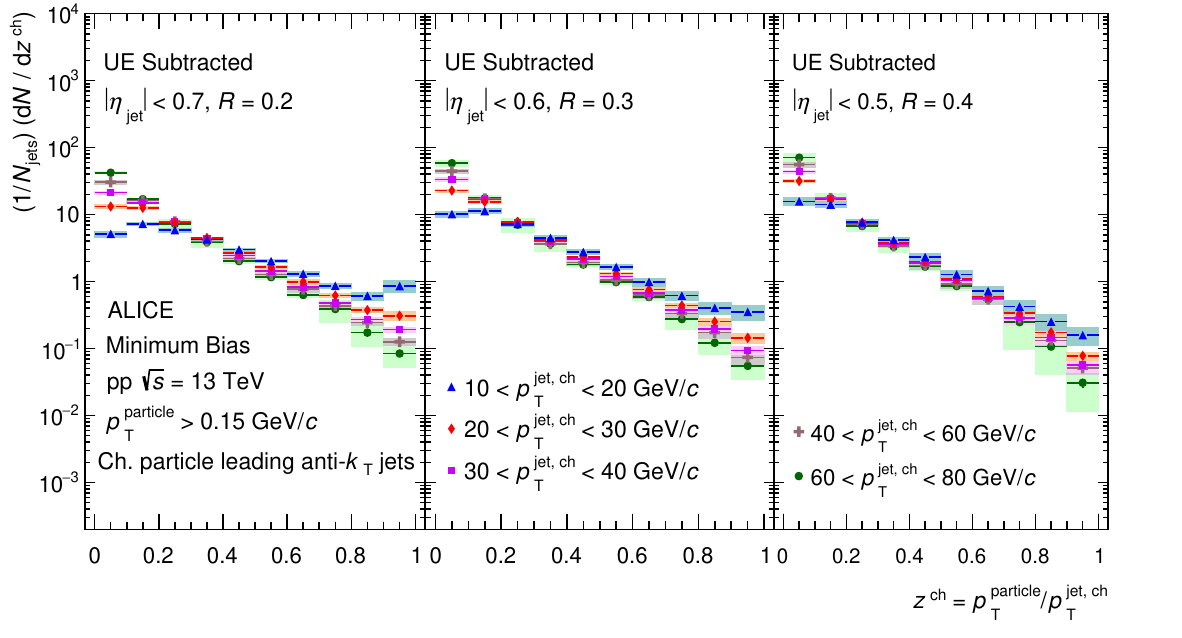}
  \includegraphics[scale=0.8]{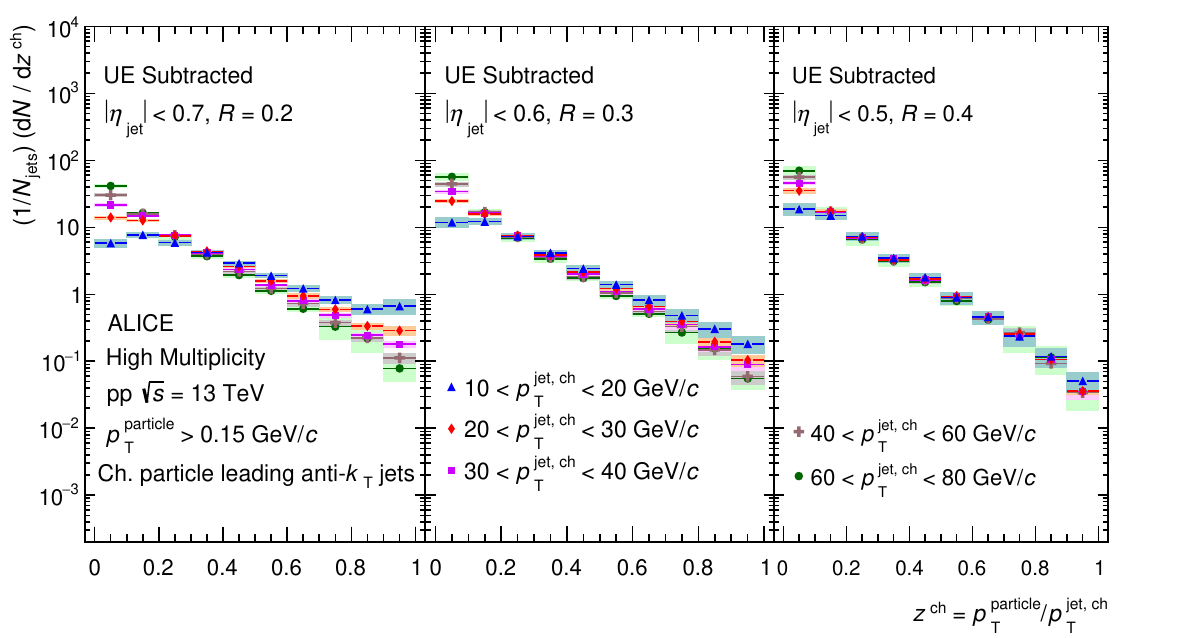}
  \caption{$z^{\rm ch}$ distributions in leading jets for different jet transverse momenta in MB (top) and HM (bottom) events for jet $R$ = 0.2 (left), 0.3 (middle), and 0.4 (right).}
  \label{FzUESubFinalAllPt_AllR_MBandHM}
\end{figure}

Figure~\ref{FzUESubFinalAllPt_AllR_MBandHM} shows the distributions of jet fragmentation function variable $z^{\rm ch}$ for jet radii 0.2 (left), 0.3 (middle), and 0.4 (right) within the jet-\pt intervals 10--20\,GeV/$c$, 20--30\,GeV/$c$, 30--40\,GeV/$c$, 40--60\,GeV/$c$, and 60--80\,GeV/$c$ for both MB (top) and HM (bottom) events. The solid markers represent the corrected results in the different jet-\pt intervals and the shaded bands are the corresponding systematic uncertainties. The statistical uncertainties are represented by vertical error bars (mostly smaller than the marker size). The distributions in different jet-\pt intervals are consistent within systematic uncertainties for wider jets ($R$ = 0.4) in HM (MB) events, in the range $0.1 < z^{\rm ch} < 1$ $(0.1 < z^{\rm ch} < 0.9)$, indicating jet-\pt independent fragmentation function. However, for narrower jets ($R$ = 0.2 and 0.3), the fragmentation functions depend on jet \pt in both MB and HM events.

\begin{figure}[h!]
  \centering
  \includegraphics[scale=0.8]{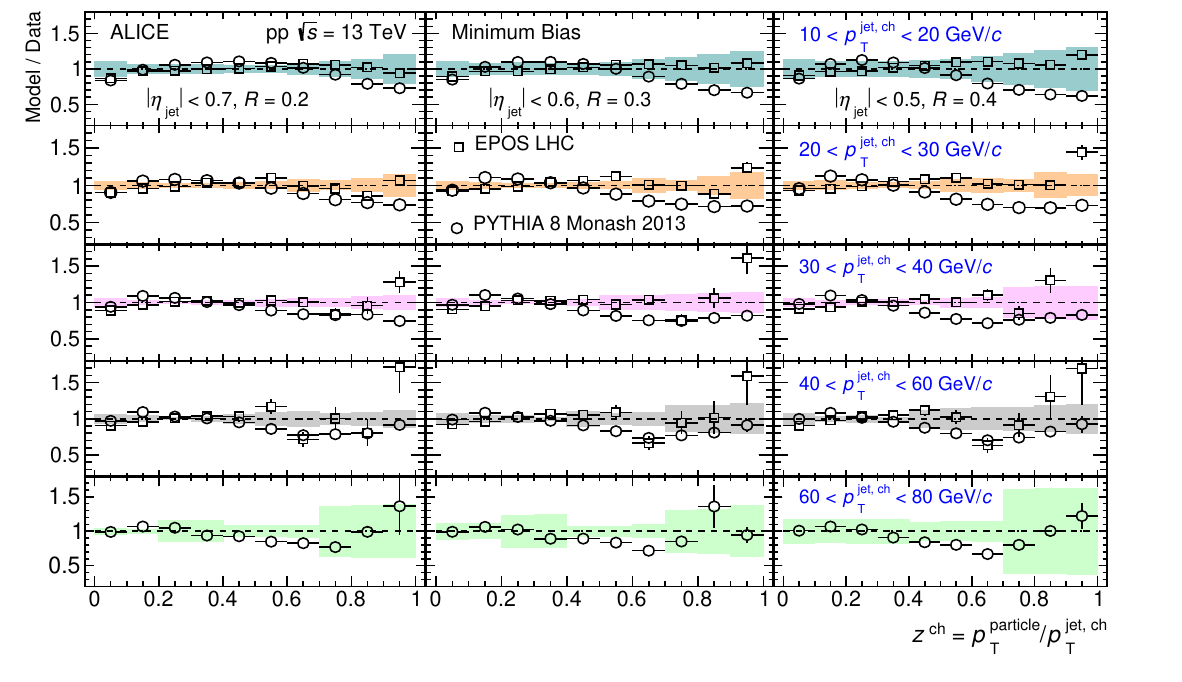}
  \includegraphics[scale=0.8]{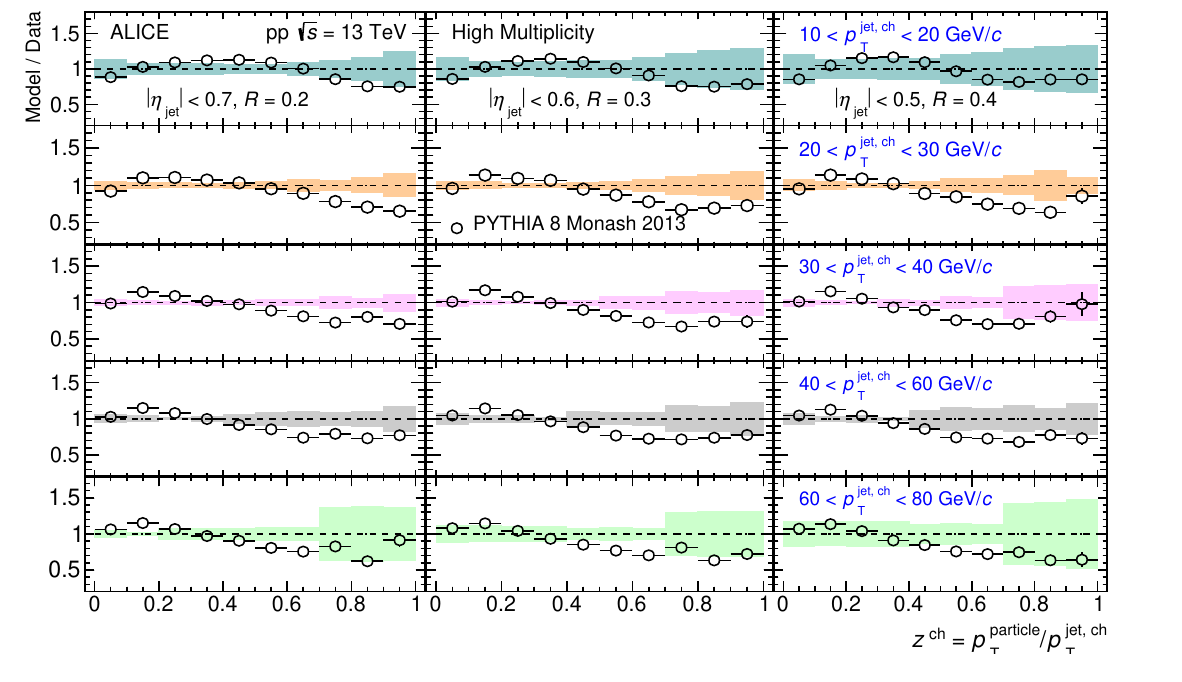}
  \caption{Top: Ratios of PYTHIA\,8 and EPOS LHC predictions to data for $z^{\rm ch}$ distributions in different $p_{\rm T}^{\rm jet,\,ch}$ intervals in MB events for jet $R$ = 0.2 (left), 0.3 (middle), and 0.4 (right). Bottom: Ratios of PYTHIA\,8 predictions to data for $z^{\rm ch}$ distributions in different $p_{\rm T}^{\rm jet,\,ch}$ intervals in HM events for jet $R$ = 0.2 (left), 0.3 (middle), and 0.4 (right).}
  \label{FzUESubFinalAllPt_AllR_MCbyData_MB}
\end{figure}

In Fig.~\ref{FzUESubFinalAllPt_AllR_MCbyData_MB}, the measured fragmentation functions are compared to predictions obtained from PYTHIA\,8 and EPOS LHC event generators for MB events (top) and with PYTHIA\,8 predictions for HM events (bottom).
For MB events, in the lowest and highest jet-\pt intervals (10--20 and 60--80 GeV/$c$), PYTHIA\,8 describes the data within systematic uncertainties; however, it underestimates the data in the intermediate jet-\pt intervals (20--30, 30--40, and 40--60 GeV/$c$) and intermediate $z^{\rm ch}$ values ($0.5 < z^{\rm ch} < 0.7$). EPOS LHC, on the other hand, reproduces the data better compared to PYTHIA\,8 for the jet-\pt intervals 10--20, 20--30, 30--40, and 40--60 GeV/c. For HM events, the ratios between the PYTHIA\,8 predictions and data in the measured jet-\pt intervals for all the jet $R$ show similar trends as observed in MB results.

\begin{figure}[h!]
  \centering
  \includegraphics[scale=0.65]{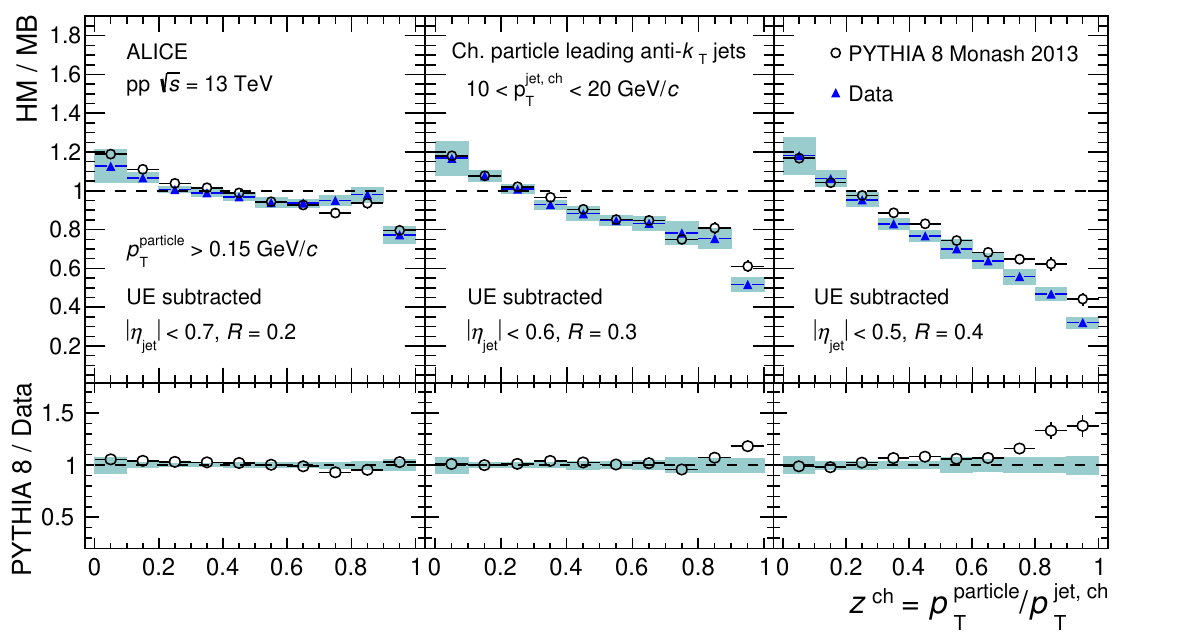}
  \includegraphics[scale=0.65]{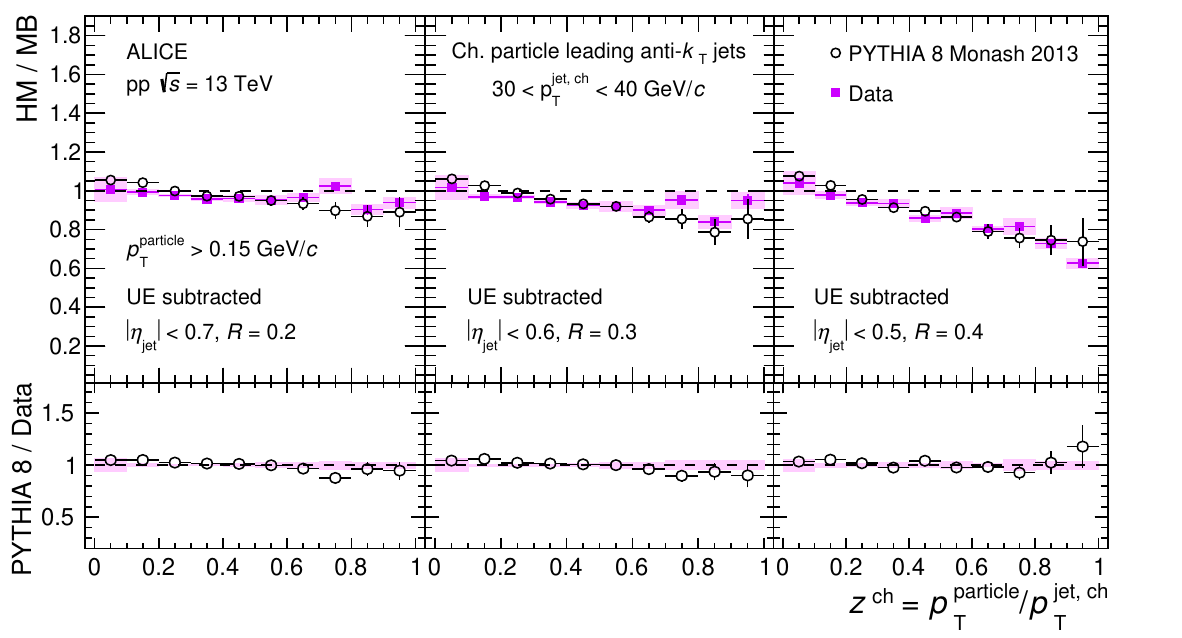}
  \includegraphics[scale=0.65]{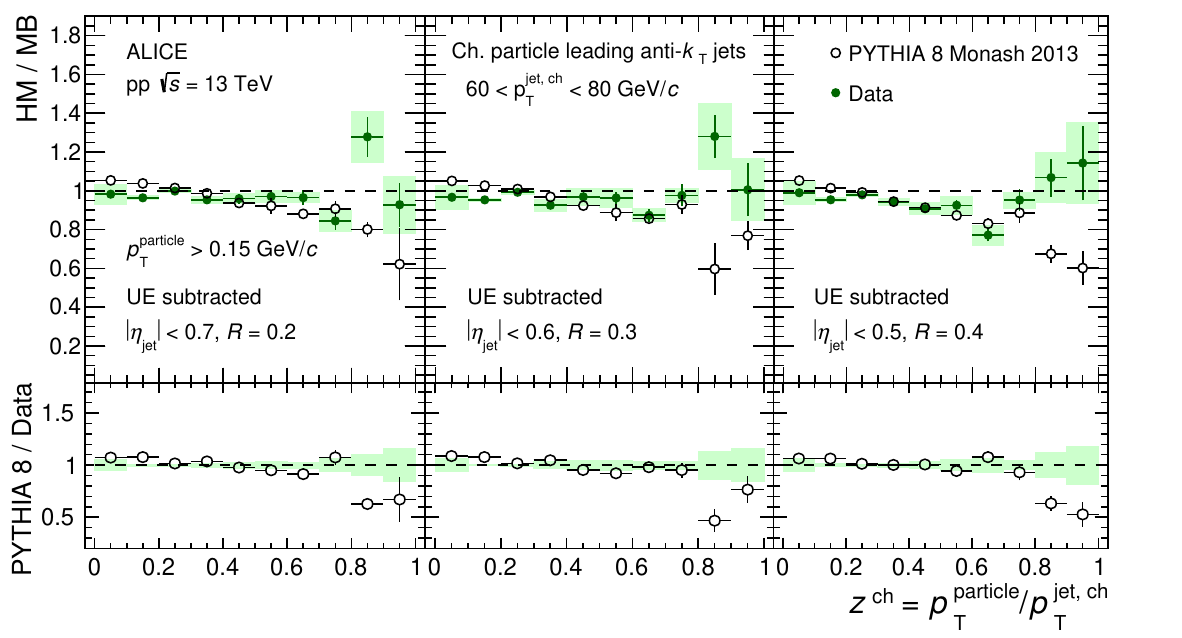}
  \caption{The ratio between HM and MB distributions of $z^{\rm ch}$ for $p_{\rm T}^{\rm jet,\,ch}$ intervals 10--20 GeV/$c$ (top), 30--40 GeV/$c$
    (middle), and 60--80 GeV/$c$ (bottom) for jet $R$ = 0.2 (left), 0.3 (middle), and 0.4 (right).}
  \label{FzUESubFinalAllPt_AllR_HMbyMB_AlljetpT}
\end{figure}

Figure~\ref{FzUESubFinalAllPt_AllR_HMbyMB_AlljetpT} depicts the ratios of $z^{\rm ch}$ distributions between HM and MB events for three jet-\pt intervals, 10--20 GeV/$c$ (top), 30--40 GeV/$c$ (middle), and 60--80 GeV/$c$ (bottom) and for jet $R$ = 0.2 (left),  0.3 (middle), and 0.4 (right). Comparisons with the PYTHIA\,8 predictions (denoted by open markers) are also shown. The distribution of $z^{\rm ch}$ in HM events is noticeably different from that from MB events for low-\pt jets (10--20 GeV/$c$), as shown in the ratio plots in the top panels of Fig.~\ref{FzUESubFinalAllPt_AllR_HMbyMB_AlljetpT}. The fragmentation probability of particles at low (high) $z^{\rm ch}$ is found to be enhanced (suppressed) in HM events compared to that in MB. This effect becomes more pronounced with increasing jet radius at a given jet \pt. The trend becomes less pronounced at higher jet \pt as it can be seen in the middle and bottom panels of Fig.~\ref{FzUESubFinalAllPt_AllR_HMbyMB_AlljetpT}. While PYTHIA\,8 shows quantitative differences from data toward higher $z^{\rm ch}$ ($>$ 0.7) values, the trends are qualitatively described except for jet $R$ = 0.4 at jet \pt = 60--80 GeV/$c$, where the statistical and systematic uncertainties are large.

A recent ALICE measurement of semi-inclusive azimuthal distributions of charged-particle jets recoiling from a high-\pt hadron trigger also shows significant azimuthal broadening in HM events compared to those in MB events and PYTHIA\,8 follows a similar broadening~\cite{FilipPaper}. A detailed investigation revealed that the HM event selection based on the V0 detector at forward rapidity introduces a bias towards multi-jet topologies, thereby affecting the azimuthal distribution. However, the measurement of intra-jet properties may evade the complication of multi-jet bias since it focuses on modifications within the leading jet, which, to first order, is independent of other jets in the event. By measuring intra-jet properties rather than jet correlations, the results shown in Fig.~\ref{FzUESubFinalAllPt_AllR_HMbyMB_AlljetpT}, therefore, provide complementary constraints on jet modification in small systems. A further investigation using less biased HM events (selected based on the total charged-particle multiplicity) in PYTHIA\,8 shows a similar modification of the jet fragmentation function variable $z^{\rm ch}$, hinting towards possible sources other than QGP formation, that may contribute to the observed modification.

From a theoretical perspective, several efforts~\cite{JetModpp1MPI, JetModpp2Rope, JetModpp3CRMPI} have been made to understand the jet modification in high-multiplicity events compared to minimum-bias ones in pp collisions. In Ref.~\cite{JetModpp3CRMPI}, a modification of jet properties in HM compared to MB events is predicted in pp collisions due to phenomena such as multiparton interactions (MPI) with color reconnection (CR) in PYTHIA\,8 as well as enhancement in the number of gluon-initiated jets in high-multiplicity events compared to that in minimum-bias collisions. 

\begin{figure}[h!]
  \centering
  \includegraphics[scale=0.39]{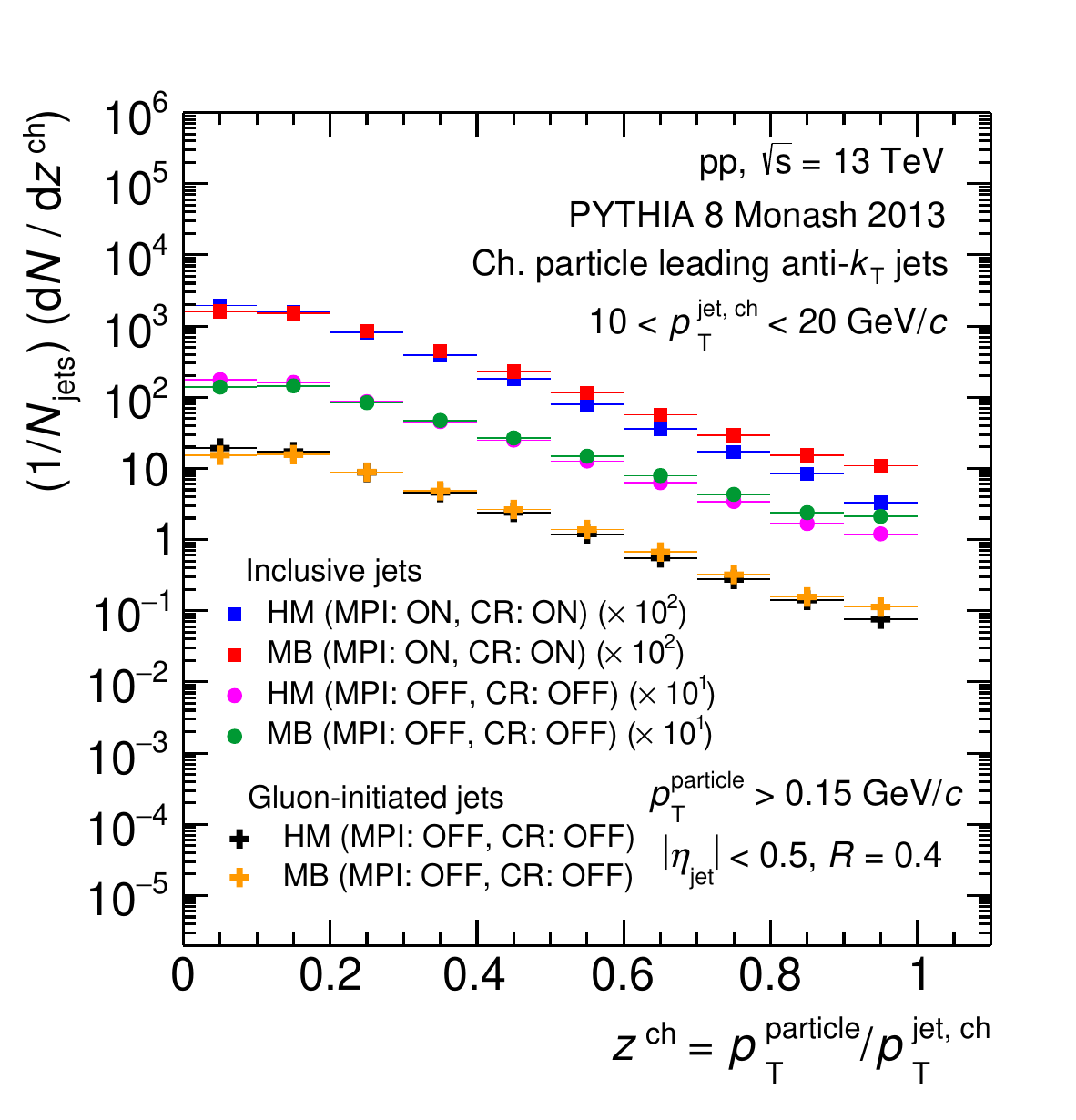}
  \includegraphics[scale=0.39]{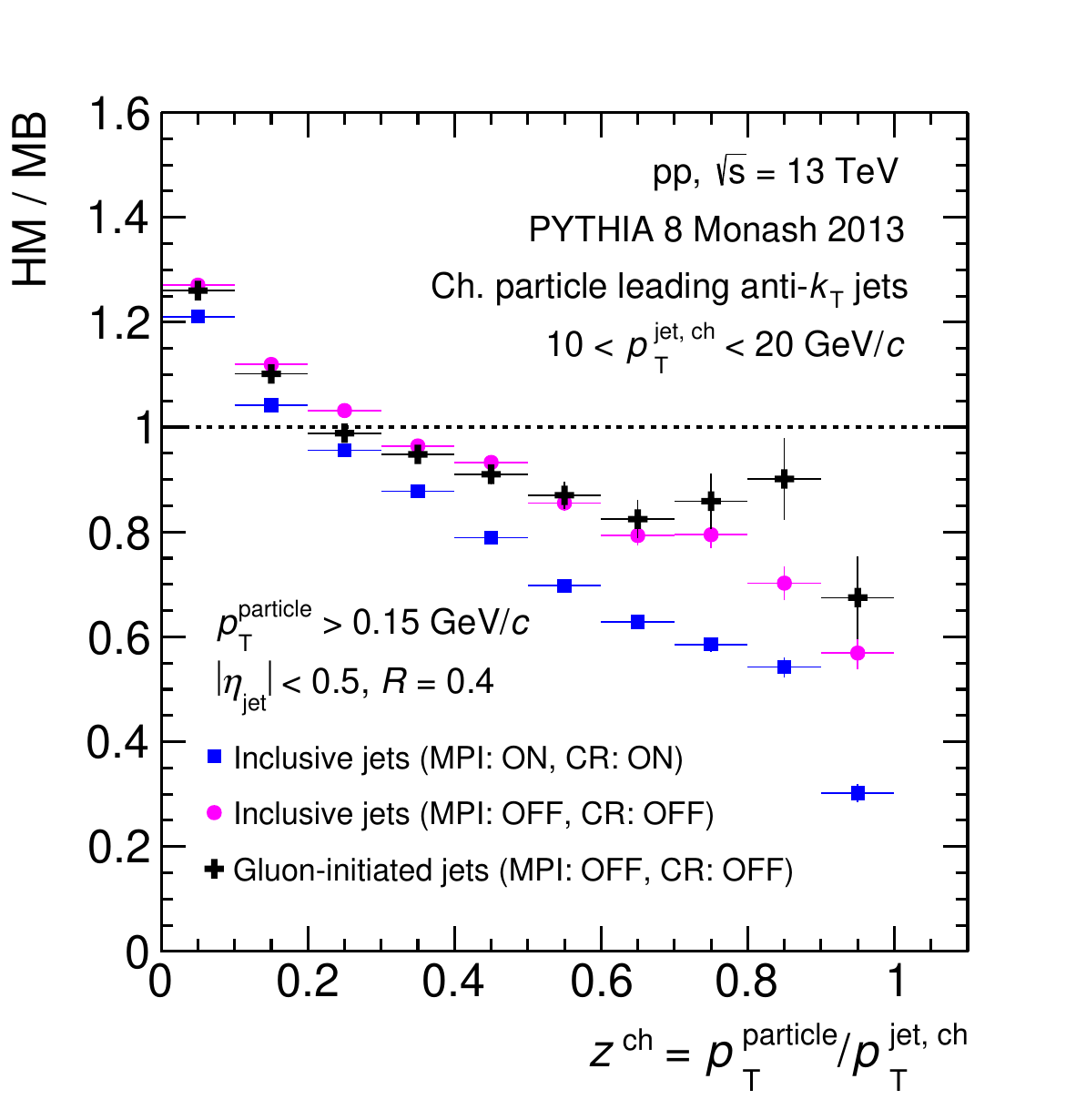}
  \caption{Left panel: Distributions of $z^{\rm ch}$  for the jet-\pt interval 10--20\,GeV/$c$  for inclusive (quark- and gluon-initiated) jets with `MPI:~ON, CR:~ON' and `MPI:~OFF, CR:~OFF' configurations, and for gluon-initiated jets with `MPI:~OFF, CR:~OFF' configuration using PYTHIA\,8. Right panel: Ratio of $z^{\rm ch}$ distributions between HM and MB events.}
  \label{Py8JetModInc}
\end{figure}

Using similar conditions for selecting MB and HM events and other kinematic selections as applied to data, the observed behavior in the ratio of $z^{\rm ch}$ distributions between HM and MB events in PYTHIA\,8 is further investigated for 10~$<p_{\rm T}^{\rm jet,\,ch}<$~20~GeV/$c$ and jet radius 0.4. Two event samples with configurations `MPI:~ON, CR:~ON' (default setting in PYTHIA\,8) and  `MPI:~OFF, CR:~OFF' (where both MPI and CR are switched off) are generated using PYTHIA\,8 for minimum-bias and high-multiplicity pp collisions at \s = 13 TeV. Figure~\ref{Py8JetModInc} (left) shows the $z^{\rm ch}$ distributions for inclusive (quark- and gluon-initiated) and gluon-initiated leading charged-particle jets in the interval 10~$<p_{\rm T}^{\rm jet,\,ch}<$~20~GeV/$c$ for both HM and MB events for the above-mentioned configurations. The ratio of $z^{\rm ch}$ distributions between HM and MB events (right panel) shows a significant modification of jet fragmentation in the presence of MPI with CR and the magnitude of the modification gets reduced when MPI and CR are switched off, indicating the dependence of jet modification on MPI and CR. The origin of the residual amount of modification in the absence of both MPI and CR is further investigated using gluon-initiated jets to check the dependence of jet modification on the nature of the initiating parton. A geometrical matching procedure based on the closest-distance approach (as applied in Ref.~\cite{JetModpp3CRMPI}) is followed to match hard-scattered partons with the leading jets. The fraction of gluon-initiated jets is found to be larger in HM events ($\sim$83\%) than in MB events ($\sim$77\%). Figure~\ref{Py8JetModInc} (right) also shows the ratio of $z^{\rm ch}$ distributions for gluon-initiated leading charged-particle jets between HM and MB events for `MPI:~OFF, CR:~OFF' configuration, showing a further, even though small, reduction of the modification with increasing multiplicity as compared to the case of inclusive jets.
These observations indicate that MPI with CR is playing major roles in the change of jet fragmentation in high-multiplicity events compared to minimum-bias events.

    \subsubsection{$\xi^{\rm ch}$ distributions}
    
    \begin{figure}[h!]
  \centering
  \includegraphics[scale=0.8]{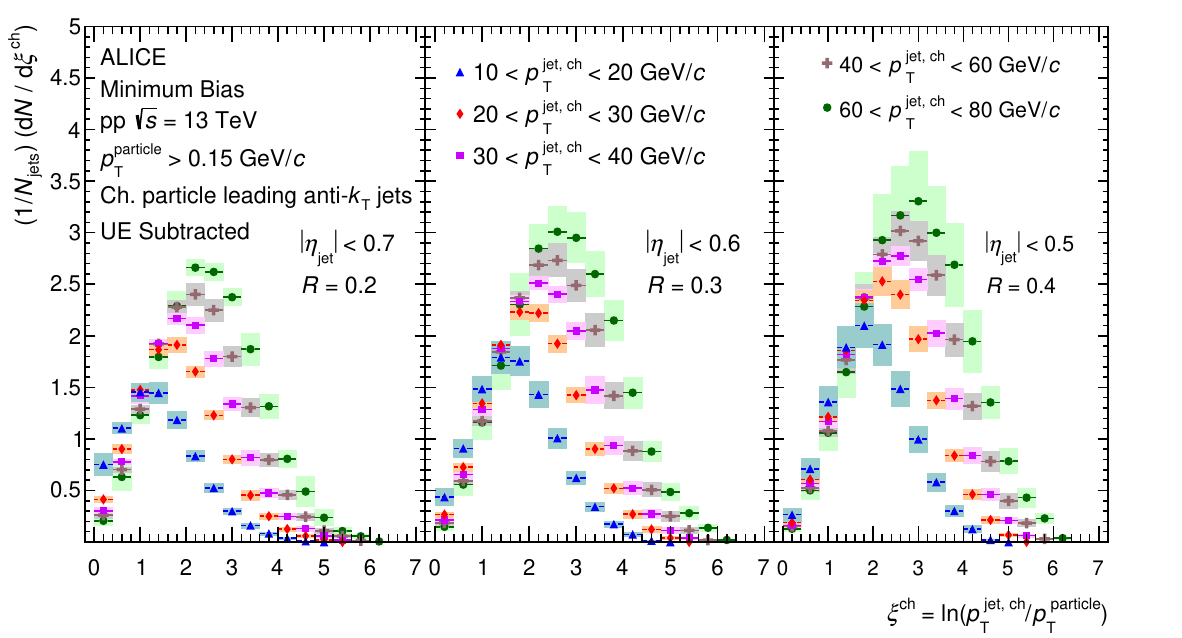}
  \includegraphics[scale=0.8]{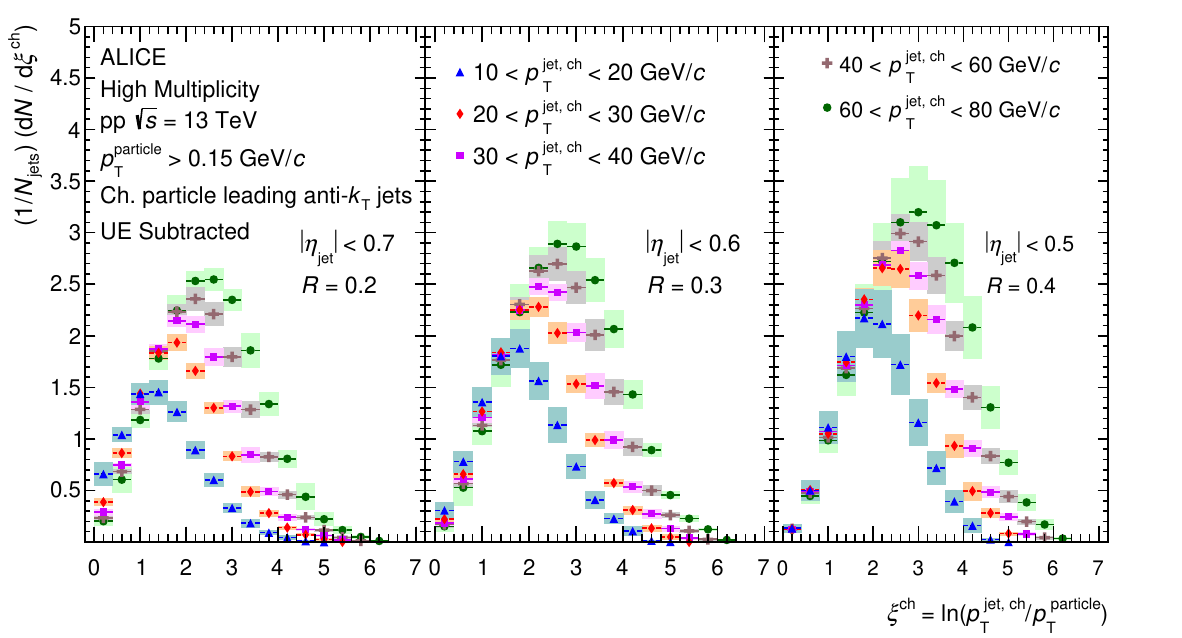}
  \caption{$\xi^{\rm ch}$ distributions in leading jets for different jet transverse momenta in MB (top) and HM (bottom) events for jet $R$ = 0.2 (left), 0.3 (middle), and 0.4 (right).}
  \label{FKsiUESubFinalAllPt_AllR_MBandHM}
\end{figure}

\begin{figure}[h!]
  \centering
  \includegraphics[scale=0.8]{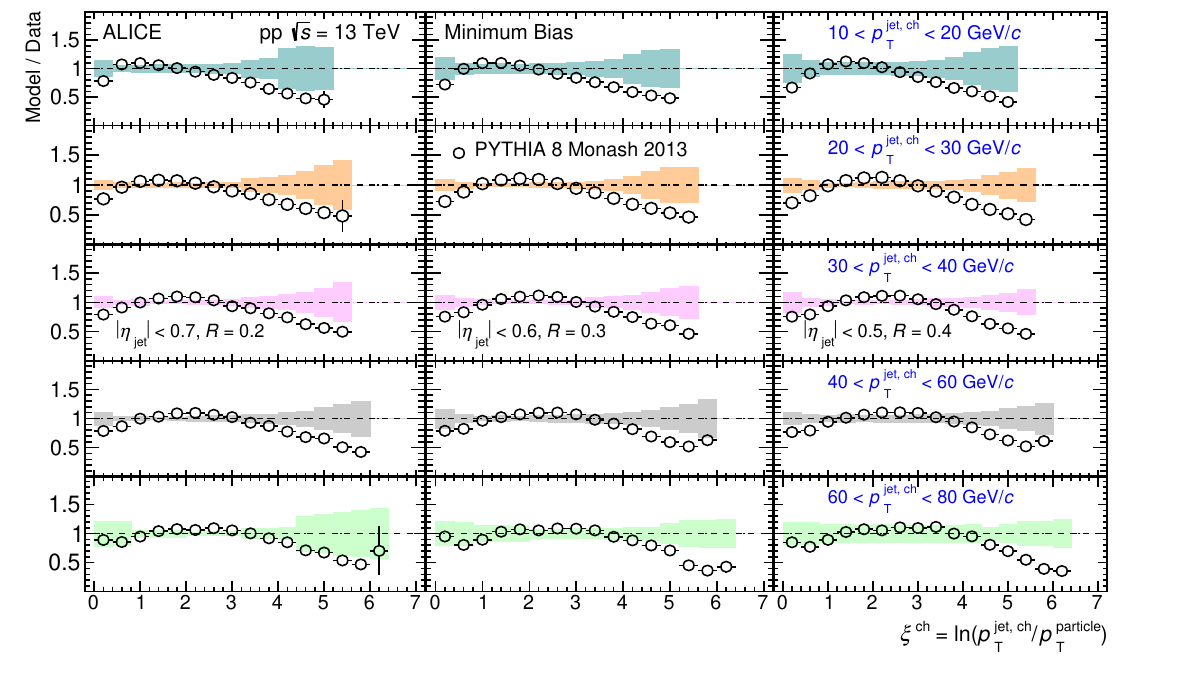}
  \includegraphics[scale=0.8]{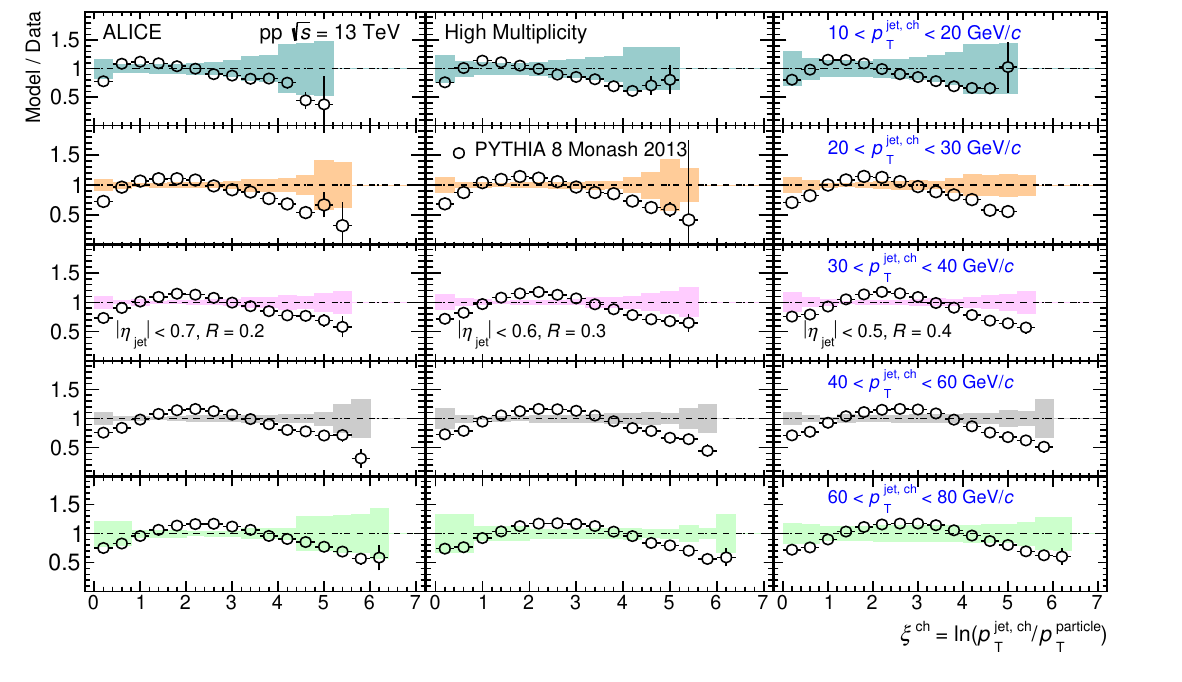}
  \caption{Ratios of PYTHIA\,8 predictions to data for $\xi^{\rm ch}$ distributions in different $p_{\rm T}^{\rm jet,\,ch}$ intervals in MB (top) and HM (bottom) events for jet $R$ = 0.2 (left), 0.3 (middle), and 0.4 (right).}
  \label{FKsiUESubFinalAllPt_AllR_MCbyData_MBandMB}
\end{figure}

\begin{figure}[h!]
  \centering
  \includegraphics[scale=0.65]{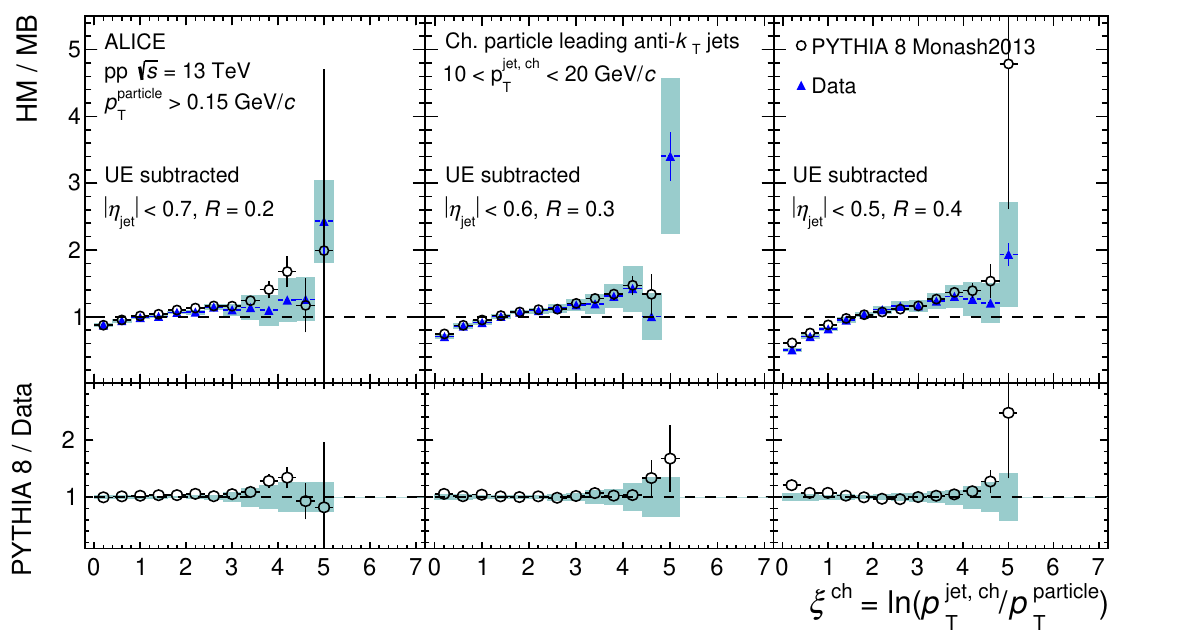}
  \includegraphics[scale=0.65]{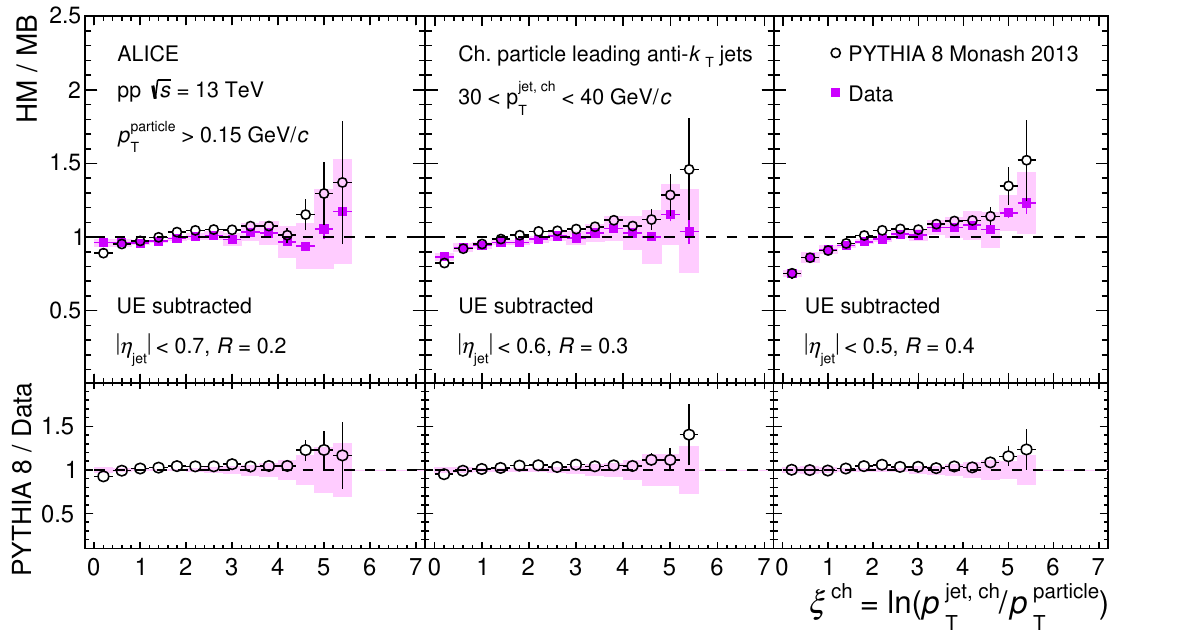}
  \includegraphics[scale=0.65]{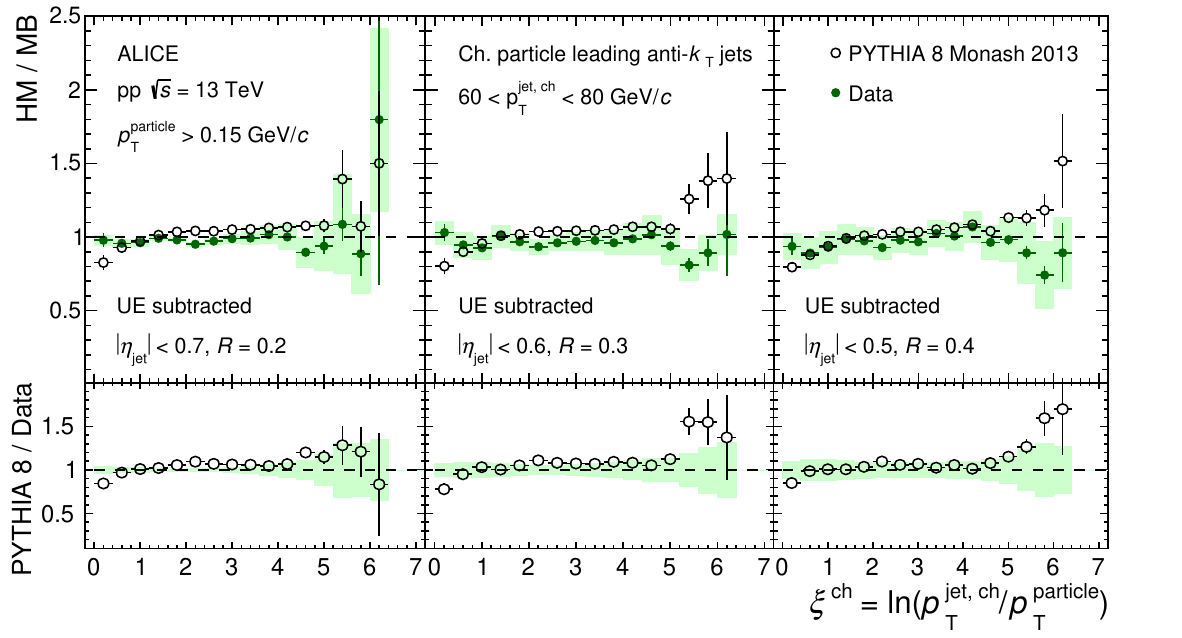}
  \caption{The ratio between HM and MB distributions of $\xi^{\rm ch}$ for $p_{\rm T}^{\rm jet,\,ch}$ intervals 10--20 GeV/$c$ (top), 30--40 GeV/$c$
  	(middle), and 60--80 GeV/$c$ (bottom) for jet $R$ = 0.2 (left), 0.3 (middle), and 0.4 (right).}
  \label{FKsiUESubFinalAllPt_AllR_HMbyMB}
\end{figure}
Figure~\ref{FKsiUESubFinalAllPt_AllR_MBandHM} shows the distributions of jet fragmentation function variable $\xi^{\rm ch}$ for jet $R$ = 0.2 (left), 0.3 (middle), and 0.4 (right) within the jet-\pt intervals 10--20, 20--30, 30--40, 40--60, and 60--80\,GeV/$c$ for both MB (top) and HM (bottom) events.
The solid markers represent the corrected results in different jet-\pt intervals and the shaded bands are the corresponding systematic uncertainties. The statistical uncertainties are represented by vertical error bars (mostly smaller than the marker size). The $\xi^{\rm ch}$ distributions highlight the low- $z^{\rm ch}$ trends in great detail. Jet-\pt independent $\xi^{\rm ch}$ distributions are observed for $\xi^{\rm ch} <$ 2 and jet $R$ = 0.4 in both MB and HM events, while the $\xi^{\rm ch}$ distributions are found to depend on jet \pt
for jet $R$ = 0.2 and 0.3. These observations are complementary to those observed in $z^{\rm ch}$ distributions. 
In addition, a pronounced peak structure, commonly known as a ``hump-backed plateau'' is observed, resulting from the suppression of low-\pt particle production predicted by QCD coherence~\cite{QCDcoherence3, QCDcoherence4, QCDcoherence1, QCDcoherence2, QCDcoherence5, QCDcoherence6}. With increasing jet \pt and rising jet $R$, the area of the $\xi^{\rm ch}$ distributions increases, complementing the results obtained from $\langle N_{\rm ch}\rangle$, indicating an increase of charged-particle multiplicity in jets with increasing jet \pt. These results show similar trends as the previous ALICE measurement in pp collisions at \s = 7~TeV~\cite{ALICExsec1}.

The comparisons of $\xi^{\rm ch}$ distributions with PYTHIA\,8 predictions are shown in Fig.~\ref{FKsiUESubFinalAllPt_AllR_MCbyData_MBandMB}. The width of the $\xi^{\rm ch}$ distributions predicted by PYTHIA\,8 is slightly smaller compared to data, whereas PYTHIA\,8 fails to reproduce the peak position of the $\xi^{\rm ch}$ distributions in data. This results in a non-flat shape in the MC-data ratios for both MB and HM events. Figure~\ref{FKsiUESubFinalAllPt_AllR_HMbyMB} shows the ratio of $\xi^{\rm ch}$ distributions between HM and MB events for three jet-\pt ranges, 10--20 GeV/$c$ (top), 30--40 GeV/$c$ (middle), and 60--80 GeV/$c$ (bottom) and for three jet $R$ = 0.2 (left),  0.3 (middle), and 0.4 (right). A clear suppression of $\xi^{\rm ch}$ distribution at low-$\xi^{\rm ch}$ values is observed in HM events compared to MB events in the lowest (10--20 GeV/$c$) jet-\pt interval for $R$ = 0.4. The amount of this suppression gets reduced with decreasing jet $R$ at a fixed jet \pt and decreases with increasing jet \pt at a given jet radius. These observations are complementary to the results as a function of $z^{\rm ch}$ reported above and the trends are well reproduced by PYTHIA\,8.

\section{Summary}
\label{summary}
In summary, this work reports the measurement of multiplicity-dependent charged-particle intra-jet properties of leading jets in pp collisions at \s = 13\,\TeV using the ALICE detector at the LHC. The mean charged-particle multiplicity $\langle N_{\rm ch}\rangle$ and jet fragmentation function variables $z^{\rm ch}$ and $\xi^{\rm ch}$ for leading jets are measured for minimum-bias and high-multiplicity events using the anti-$k_{\rm T}$ jet finding algorithm with $R$ = 0.2, 0.3, and 0.4. A monotonic increase in $\langle N_{\rm ch}\rangle$ is observed in both MB and HM events as a function of jet \pt as well as with increasing jet radius R. It is found to be slightly larger in high-multiplicity events in comparison with minimum-bias ones; PYTHIA\,8 also exhibits a similar pattern. A jet-\pt independent jet fragmentation is observed in both MB and HM events within certain ranges of $z^{\rm ch}$ and $\xi^{\rm ch}$ values only for wider jets ($R$ = 0.4). EPOS LHC reproduces the $z^{\rm ch}$ distributions better than PYTHIA\,8 in MB events. The observed ``hump-backed plateau'' structure in $\xi^{\rm ch}$ distributions originates from the suppression of low-\pt particle production predicted by QCD coherence. The $\xi^{\rm ch}$ distributions for both MB and HM events are qualitatively reproduced by PYTHIA\,8.
\par
The fragmentation functions in HM events are noticeably different from those in MB events. The probability of jet fragmentation into particles with low $z^{\rm ch}$ gets enhanced, followed by a suppression of high-$z^{\rm ch}$ particles in HM events compared to that in MB. The observed jet modification is more prominent for low-\pt jets (10--20 GeV/c) with larger jet radius ($R$ = 0.4) and is reduced with increasing jet \pt at a given radius. These trends are qualitatively reproduced by PYTHIA\,8. Similar conclusions are obtained when studying the fragmentation function in MB and HM events using the $\xi^{\rm ch}$ variable.
\par
Signatures of collectivity, previously associated with QGP production, have been observed in event multiplicity-dependent measurements in the soft sector of QCD. Recently, a selection bias towards multi-jet topology has been argued to affect the observed azimuthal broadening in the sample of high-multiplicity events defined from the V0M signal amplitudes. The modifications of intra-jet properties for leading jets are independent of the presence of other jets in an event and are therefore less prone to such biases. The multiplicity-dependent measurements of intra-jet properties presented in this article show
that increasing event multiplicity in small collision systems also influences hard probes such as jets. An investigation using PYTHIA\,8 with a less biased HM event selection also shows a similar amount of modification. A detailed study using PYTHIA\,8 shows that the major source of the modification in jet fragmentation is multiparton interactions with color reconnection in HM events. One can conclude from here that jet modification is observed in small systems with increasing multiplicity, shifting the question towards how one can attribute the observed modification to different causes, e.g., multiparton interactions, jet bias from HM event selection or jet quenching in mini-QGP. Since PYTHIA\,8 captures most of the features of the data, the measured modifications cannot be directly interpreted as due to the formation of a QGP in high-multiplicity pp collisions. The measurements of intra-jet properties reported in this work provide new constraints to mechanisms underlying jet modification in small systems. More precise measurements of these observables could provide better insights into the jet modification in small collision systems with high final-state multiplicity.


\newenvironment{acknowledgement}{\relax}{\relax}
\begin{acknowledgement}
\section*{Acknowledgements}

The ALICE Collaboration would like to thank all its engineers and technicians for their invaluable contributions to the construction of the experiment and the CERN accelerator teams for the outstanding performance of the LHC complex.
The ALICE Collaboration gratefully acknowledges the resources and support provided by all Grid centres and the Worldwide LHC Computing Grid (WLCG) collaboration.
The ALICE Collaboration acknowledges the following funding agencies for their support in building and running the ALICE detector:
A. I. Alikhanyan National Science Laboratory (Yerevan Physics Institute) Foundation (ANSL), State Committee of Science and World Federation of Scientists (WFS), Armenia;
Austrian Academy of Sciences, Austrian Science Fund (FWF): [M 2467-N36] and Nationalstiftung f\"{u}r Forschung, Technologie und Entwicklung, Austria;
Ministry of Communications and High Technologies, National Nuclear Research Center, Azerbaijan;
Conselho Nacional de Desenvolvimento Cient\'{\i}fico e Tecnol\'{o}gico (CNPq), Financiadora de Estudos e Projetos (Finep), Funda\c{c}\~{a}o de Amparo \`{a} Pesquisa do Estado de S\~{a}o Paulo (FAPESP) and Universidade Federal do Rio Grande do Sul (UFRGS), Brazil;
Bulgarian Ministry of Education and Science, within the National Roadmap for Research Infrastructures 2020-2027 (object CERN), Bulgaria;
Ministry of Education of China (MOEC) , Ministry of Science \& Technology of China (MSTC) and National Natural Science Foundation of China (NSFC), China;
Ministry of Science and Education and Croatian Science Foundation, Croatia;
Centro de Aplicaciones Tecnol\'{o}gicas y Desarrollo Nuclear (CEADEN), Cubaenerg\'{\i}a, Cuba;
Ministry of Education, Youth and Sports of the Czech Republic, Czech Republic;
The Danish Council for Independent Research | Natural Sciences, the VILLUM FONDEN and Danish National Research Foundation (DNRF), Denmark;
Helsinki Institute of Physics (HIP), Finland;
Commissariat \`{a} l'Energie Atomique (CEA) and Institut National de Physique Nucl\'{e}aire et de Physique des Particules (IN2P3) and Centre National de la Recherche Scientifique (CNRS), France;
Bundesministerium f\"{u}r Bildung und Forschung (BMBF) and GSI Helmholtzzentrum f\"{u}r Schwerionenforschung GmbH, Germany;
General Secretariat for Research and Technology, Ministry of Education, Research and Religions, Greece;
National Research, Development and Innovation Office, Hungary;
Department of Atomic Energy Government of India (DAE), Department of Science and Technology, Government of India (DST), University Grants Commission, Government of India (UGC) and Council of Scientific and Industrial Research (CSIR), India;
National Research and Innovation Agency - BRIN, Indonesia;
Istituto Nazionale di Fisica Nucleare (INFN), Italy;
Japanese Ministry of Education, Culture, Sports, Science and Technology (MEXT) and Japan Society for the Promotion of Science (JSPS) KAKENHI, Japan;
Consejo Nacional de Ciencia (CONACYT) y Tecnolog\'{i}a, through Fondo de Cooperaci\'{o}n Internacional en Ciencia y Tecnolog\'{i}a (FONCICYT) and Direcci\'{o}n General de Asuntos del Personal Academico (DGAPA), Mexico;
Nederlandse Organisatie voor Wetenschappelijk Onderzoek (NWO), Netherlands;
The Research Council of Norway, Norway;
Commission on Science and Technology for Sustainable Development in the South (COMSATS), Pakistan;
Pontificia Universidad Cat\'{o}lica del Per\'{u}, Peru;
Ministry of Education and Science, National Science Centre and WUT ID-UB, Poland;
Korea Institute of Science and Technology Information and National Research Foundation of Korea (NRF), Republic of Korea;
Ministry of Education and Scientific Research, Institute of Atomic Physics, Ministry of Research and Innovation and Institute of Atomic Physics and Universitatea Nationala de Stiinta si Tehnologie Politehnica Bucuresti, Romania;
Ministry of Education, Science, Research and Sport of the Slovak Republic, Slovakia;
National Research Foundation of South Africa, South Africa;
Swedish Research Council (VR) and Knut \& Alice Wallenberg Foundation (KAW), Sweden;
European Organization for Nuclear Research, Switzerland;
Suranaree University of Technology (SUT), National Science and Technology Development Agency (NSTDA) and National Science, Research and Innovation Fund (NSRF via PMU-B B05F650021), Thailand;
Turkish Energy, Nuclear and Mineral Research Agency (TENMAK), Turkey;
National Academy of  Sciences of Ukraine, Ukraine;
Science and Technology Facilities Council (STFC), United Kingdom;
National Science Foundation of the United States of America (NSF) and United States Department of Energy, Office of Nuclear Physics (DOE NP), United States of America.
In addition, individual groups or members have received support from:
Czech Science Foundation (grant no. 23-07499S), Czech Republic;
European Research Council, Strong 2020 - Horizon 2020 (grant nos. 950692, 824093), European Union;
ICSC - Centro Nazionale di Ricerca in High Performance Computing, Big Data and Quantum Computing, European Union - NextGenerationEU;
Academy of Finland (Center of Excellence in Quark Matter) (grant nos. 346327, 346328), Finland.

\end{acknowledgement}

\bibliographystyle{utphys}   
\bibliography{bibliography}
\newpage
\appendix
%
%

\section{The ALICE Collaboration}
\label{app:collab}
\begin{flushleft} 
\small

S.~Acharya\,\orcidlink{0000-0002-9213-5329}\,$^{\rm 128}$, 
D.~Adamov\'{a}\,\orcidlink{0000-0002-0504-7428}\,$^{\rm 87}$, 
G.~Aglieri Rinella\,\orcidlink{0000-0002-9611-3696}\,$^{\rm 33}$, 
L.~Aglietta$^{\rm 25}$, 
M.~Agnello\,\orcidlink{0000-0002-0760-5075}\,$^{\rm 30}$, 
N.~Agrawal\,\orcidlink{0000-0003-0348-9836}\,$^{\rm 52}$, 
Z.~Ahammed\,\orcidlink{0000-0001-5241-7412}\,$^{\rm 136}$, 
S.~Ahmad\,\orcidlink{0000-0003-0497-5705}\,$^{\rm 16}$, 
S.U.~Ahn\,\orcidlink{0000-0001-8847-489X}\,$^{\rm 72}$, 
I.~Ahuja\,\orcidlink{0000-0002-4417-1392}\,$^{\rm 38}$
% ...

\section*{Affiliation Notes}

$^{\rm I}$ Deceased\\
$^{\rm II}$ Also at: Max-Planck-Institut fur Physik, Munich, Germany\\
$^{\rm III}$ Also at: Italian National Agency for New Technologies, Energy and Sustainable Economic Development (ENEA), Bologna, Italy\\
$^{\rm IV}$ Also at: Dipartimento DET del Politecnico di Torino, Turin, Italy\\
$^{\rm V}$ Also at: Yildiz Technical University, Istanbul, T\"{u}rkiye\\
$^{\rm VI}$ Also at: Department of Applied Physics, Aligarh Muslim University, Aligarh, India\\
$^{\rm VII}$ Also at: Institute of Theoretical Physics, University of Wroclaw, Poland\\
$^{\rm VIII}$ Also at: An institution covered by a cooperation agreement with CERN\\

\section*{Collaboration Institutes}

$^{1}$ A.I. Alikhanyan National Science Laboratory (Yerevan Physics Institute) Foundation, Yerevan, Armenia\\
$^{2}$ AGH University of Krakow, Cracow, Poland\\
$^{3}$ Bogolyubov Institute for Theoretical Physics, National Academy of Sciences of Ukraine, Kiev, Ukraine\\
$^{4}$ Bose Institute, Department of Physics  and Centre for Astroparticle Physics and Space Science (CAPSS), Kolkata, India\\
$^{5}$ California Polytechnic State University, San Luis Obispo, California, United States\\
$^{6}$ Central China Normal University, Wuhan, China\\
$^{7}$ Centro de Aplicaciones Tecnol\'{o}gicas y Desarrollo Nuclear (CEADEN), Havana, Cuba\\
$^{8}$ Centro de Investigaci\'{o}n y de Estudios Avanzados (CINVESTAV), Mexico City and M\'{e}rida, Mexico\\
$^{9}$ Chicago State University, Chicago, Illinois, United States\\
$^{10}$ China Institute of Atomic Energy, Beijing, China\\
$^{11}$ China University of Geosciences, Wuhan, China\\
$^{12}$ Chungbuk National University, Cheongju, Republic of Korea\\
$^{13}$ Comenius University Bratislava, Faculty of Mathematics, Physics and Informatics, Bratislava, Slovak Republic\\
$^{14}$ COMSATS University Islamabad, Islamabad, Pakistan\\
$^{15}$ Creighton University, Omaha, Nebraska, United States\\
$^{16}$ Department of Physics, Aligarh Muslim University, Aligarh, India\\
$^{17}$ Department of Physics, Pusan National University, Pusan, Republic of Korea\\
$^{18}$ Department of Physics, Sejong University, Seoul, Republic of Korea\\
$^{19}$ Department of Physics, University of California, Berkeley, California, United States\\
$^{20}$ Department of Physics, University of Oslo, Oslo, Norway\\
$^{21}$ Department of Physics and Technology, University of Bergen, Bergen, Norway\\
$^{22}$ Dipartimento di Fisica, Universit\`{a} di Pavia, Pavia, Italy\\
$^{23}$ Dipartimento di Fisica dell'Universit\`{a} and Sezione INFN, Cagliari, Italy\\
$^{24}$ Dipartimento di Fisica dell'Universit\`{a} and Sezione INFN, Trieste, Italy\\
$^{25}$ Dipartimento di Fisica dell'Universit\`{a} and Sezione INFN, Turin, Italy\\
$^{26}$ Dipartimento di Fisica e Astronomia dell'Universit\`{a} and Sezione INFN, Bologna, Italy\\
$^{27}$ Dipartimento di Fisica e Astronomia dell'Universit\`{a} and Sezione INFN, Catania, Italy\\
$^{28}$ Dipartimento di Fisica e Astronomia dell'Universit\`{a} and Sezione INFN, Padova, Italy\\
$^{29}$ Dipartimento di Fisica `E.R.~Caianiello' dell'Universit\`{a} and Gruppo Collegato INFN, Salerno, Italy\\
$^{30}$ Dipartimento DISAT del Politecnico and Sezione INFN, Turin, Italy\\
$^{31}$ Dipartimento di Scienze MIFT, Universit\`{a} di Messina, Messina, Italy\\
$^{32}$ Dipartimento Interateneo di Fisica `M.~Merlin' and Sezione INFN, Bari, Italy\\
$^{33}$ European Organization for Nuclear Research (CERN), Geneva, Switzerland\\
$^{34}$ Faculty of Electrical Engineering, Mechanical Engineering and Naval Architecture, University of Split, Split, Croatia\\
$^{35}$ Faculty of Engineering and Science, Western Norway University of Applied Sciences, Bergen, Norway\\
$^{36}$ Faculty of Nuclear Sciences and Physical Engineering, Czech Technical University in Prague, Prague, Czech Republic\\
$^{37}$ Faculty of Physics, Sofia University, Sofia, Bulgaria\\
$^{38}$ Faculty of Science, P.J.~\v{S}af\'{a}rik University, Ko\v{s}ice, Slovak Republic\\
$^{39}$ Frankfurt Institute for Advanced Studies, Johann Wolfgang Goethe-Universit\"{a}t Frankfurt, Frankfurt, Germany\\
$^{40}$ Fudan University, Shanghai, China\\
$^{41}$ Gangneung-Wonju National University, Gangneung, Republic of Korea\\
$^{42}$ Gauhati University, Department of Physics, Guwahati, India\\
$^{43}$ Helmholtz-Institut f\"{u}r Strahlen- und Kernphysik, Rheinische Friedrich-Wilhelms-Universit\"{a}t Bonn, Bonn, Germany\\
$^{44}$ Helsinki Institute of Physics (HIP), Helsinki, Finland\\
$^{45}$ High Energy Physics Group,  Universidad Aut\'{o}noma de Puebla, Puebla, Mexico\\
$^{46}$ Horia Hulubei National Institute of Physics and Nuclear Engineering, Bucharest, Romania\\
$^{47}$ HUN-REN Wigner Research Centre for Physics, Budapest, Hungary\\
$^{48}$ Indian Institute of Technology Bombay (IIT), Mumbai, India\\
$^{49}$ Indian Institute of Technology Indore, Indore, India\\
$^{50}$ INFN, Laboratori Nazionali di Frascati, Frascati, Italy\\
$^{51}$ INFN, Sezione di Bari, Bari, Italy\\
$^{52}$ INFN, Sezione di Bologna, Bologna, Italy\\
$^{53}$ INFN, Sezione di Cagliari, Cagliari, Italy\\
$^{54}$ INFN, Sezione di Catania, Catania, Italy\\
$^{55}$ INFN, Sezione di Padova, Padova, Italy\\
$^{56}$ INFN, Sezione di Pavia, Pavia, Italy\\
$^{57}$ INFN, Sezione di Torino, Turin, Italy\\
$^{58}$ INFN, Sezione di Trieste, Trieste, Italy\\
$^{59}$ Inha University, Incheon, Republic of Korea\\
$^{60}$ Institute for Gravitational and Subatomic Physics (GRASP), Utrecht University/Nikhef, Utrecht, Netherlands\\
$^{61}$ Institute of Experimental Physics, Slovak Academy of Sciences, Ko\v{s}ice, Slovak Republic\\
$^{62}$ Institute of Physics, Homi Bhabha National Institute, Bhubaneswar, India\\
$^{63}$ Institute of Physics of the Czech Academy of Sciences, Prague, Czech Republic\\
$^{64}$ Institute of Space Science (ISS), Bucharest, Romania\\
$^{65}$ Institut f\"{u}r Kernphysik, Johann Wolfgang Goethe-Universit\"{a}t Frankfurt, Frankfurt, Germany\\
$^{66}$ Instituto de Ciencias Nucleares, Universidad Nacional Aut\'{o}noma de M\'{e}xico, Mexico City, Mexico\\
$^{67}$ Instituto de F\'{i}sica, Universidade Federal do Rio Grande do Sul (UFRGS), Porto Alegre, Brazil\\
$^{68}$ Instituto de F\'{\i}sica, Universidad Nacional Aut\'{o}noma de M\'{e}xico, Mexico City, Mexico\\
$^{69}$ iThemba LABS, National Research Foundation, Somerset West, South Africa\\
$^{70}$ Jeonbuk National University, Jeonju, Republic of Korea\\
$^{71}$ Johann-Wolfgang-Goethe Universit\"{a}t Frankfurt Institut f\"{u}r Informatik, Fachbereich Informatik und Mathematik, Frankfurt, Germany\\
$^{72}$ Korea Institute of Science and Technology Information, Daejeon, Republic of Korea\\
$^{73}$ KTO Karatay University, Konya, Turkey\\
$^{74}$ Laboratoire de Physique Subatomique et de Cosmologie, Universit\'{e} Grenoble-Alpes, CNRS-IN2P3, Grenoble, France\\
$^{75}$ Lawrence Berkeley National Laboratory, Berkeley, California, United States\\
$^{76}$ Lund University Department of Physics, Division of Particle Physics, Lund, Sweden\\
$^{77}$ Nagasaki Institute of Applied Science, Nagasaki, Japan\\
$^{78}$ Nara Women{'}s University (NWU), Nara, Japan\\
$^{79}$ National and Kapodistrian University of Athens, School of Science, Department of Physics , Athens, Greece\\
$^{80}$ National Centre for Nuclear Research, Warsaw, Poland\\
$^{81}$ National Institute of Science Education and Research, Homi Bhabha National Institute, Jatni, India\\
$^{82}$ National Nuclear Research Center, Baku, Azerbaijan\\
$^{83}$ National Research and Innovation Agency - BRIN, Jakarta, Indonesia\\
$^{84}$ Niels Bohr Institute, University of Copenhagen, Copenhagen, Denmark\\
$^{85}$ Nikhef, National institute for subatomic physics, Amsterdam, Netherlands\\
$^{86}$ Nuclear Physics Group, STFC Daresbury Laboratory, Daresbury, United Kingdom\\
$^{87}$ Nuclear Physics Institute of the Czech Academy of Sciences, Husinec-\v{R}e\v{z}, Czech Republic\\
$^{88}$ Oak Ridge National Laboratory, Oak Ridge, Tennessee, United States\\
$^{89}$ Ohio State University, Columbus, Ohio, United States\\
$^{90}$ Physics department, Faculty of science, University of Zagreb, Zagreb, Croatia\\
$^{91}$ Physics Department, Panjab University, Chandigarh, India\\
$^{92}$ Physics Department, University of Jammu, Jammu, India\\
$^{93}$ Physics Program and International Institute for Sustainability with Knotted Chiral Meta Matter (SKCM2), Hiroshima University, Hiroshima, Japan\\
$^{94}$ Physikalisches Institut, Eberhard-Karls-Universit\"{a}t T\"{u}bingen, T\"{u}bingen, Germany\\
$^{95}$ Physikalisches Institut, Ruprecht-Karls-Universit\"{a}t Heidelberg, Heidelberg, Germany\\
$^{96}$ Physik Department, Technische Universit\"{a}t M\"{u}nchen, Munich, Germany\\
$^{97}$ Politecnico di Bari and Sezione INFN, Bari, Italy\\
$^{98}$ Research Division and ExtreMe Matter Institute EMMI, GSI Helmholtzzentrum f\"ur Schwerionenforschung GmbH, Darmstadt, Germany\\
$^{99}$ Saga University, Saga, Japan\\
$^{100}$ Saha Institute of Nuclear Physics, Homi Bhabha National Institute, Kolkata, India\\
$^{101}$ School of Physics and Astronomy, University of Birmingham, Birmingham, United Kingdom\\
$^{102}$ Secci\'{o}n F\'{\i}sica, Departamento de Ciencias, Pontificia Universidad Cat\'{o}lica del Per\'{u}, Lima, Peru\\
$^{103}$ Stefan Meyer Institut f\"{u}r Subatomare Physik (SMI), Vienna, Austria\\
$^{104}$ SUBATECH, IMT Atlantique, Nantes Universit\'{e}, CNRS-IN2P3, Nantes, France\\
$^{105}$ Sungkyunkwan University, Suwon City, Republic of Korea\\
$^{106}$ Suranaree University of Technology, Nakhon Ratchasima, Thailand\\
$^{107}$ Technical University of Ko\v{s}ice, Ko\v{s}ice, Slovak Republic\\
$^{108}$ The Henryk Niewodniczanski Institute of Nuclear Physics, Polish Academy of Sciences, Cracow, Poland\\
$^{109}$ The University of Texas at Austin, Austin, Texas, United States\\
$^{110}$ Universidad Aut\'{o}noma de Sinaloa, Culiac\'{a}n, Mexico\\
$^{111}$ Universidade de S\~{a}o Paulo (USP), S\~{a}o Paulo, Brazil\\
$^{112}$ Universidade Estadual de Campinas (UNICAMP), Campinas, Brazil\\
$^{113}$ Universidade Federal do ABC, Santo Andre, Brazil\\
$^{114}$ Universitatea Nationala de Stiinta si Tehnologie Politehnica Bucuresti, Bucharest, Romania\\
$^{115}$ University of Cape Town, Cape Town, South Africa\\
$^{116}$ University of Derby, Derby, United Kingdom\\
$^{117}$ University of Houston, Houston, Texas, United States\\
$^{118}$ University of Jyv\"{a}skyl\"{a}, Jyv\"{a}skyl\"{a}, Finland\\
$^{119}$ University of Kansas, Lawrence, Kansas, United States\\
$^{120}$ University of Liverpool, Liverpool, United Kingdom\\
$^{121}$ University of Science and Technology of China, Hefei, China\\
$^{122}$ University of South-Eastern Norway, Kongsberg, Norway\\
$^{123}$ University of Tennessee, Knoxville, Tennessee, United States\\
$^{124}$ University of the Witwatersrand, Johannesburg, South Africa\\
$^{125}$ University of Tokyo, Tokyo, Japan\\
$^{126}$ University of Tsukuba, Tsukuba, Japan\\
$^{127}$ Universit\"{a}t M\"{u}nster, Institut f\"{u}r Kernphysik, M\"{u}nster, Germany\\
$^{128}$ Universit\'{e} Clermont Auvergne, CNRS/IN2P3, LPC, Clermont-Ferrand, France\\
$^{129}$ Universit\'{e} de Lyon, CNRS/IN2P3, Institut de Physique des 2 Infinis de Lyon, Lyon, France\\
$^{130}$ Universit\'{e} de Strasbourg, CNRS, IPHC UMR 7178, F-67000 Strasbourg, France, Strasbourg, France\\
$^{131}$ Universit\'{e} Paris-Saclay, Centre d'Etudes de Saclay (CEA), IRFU, D\'{e}partment de Physique Nucl\'{e}aire (DPhN), Saclay, France\\
$^{132}$ Universit\'{e}  Paris-Saclay, CNRS/IN2P3, IJCLab, Orsay, France\\
$^{133}$ Universit\`{a} degli Studi di Foggia, Foggia, Italy\\
$^{134}$ Universit\`{a} del Piemonte Orientale, Vercelli, Italy\\
$^{135}$ Universit\`{a} di Brescia, Brescia, Italy\\
$^{136}$ Variable Energy Cyclotron Centre, Homi Bhabha National Institute, Kolkata, India\\
$^{137}$ Warsaw University of Technology, Warsaw, Poland\\
$^{138}$ Wayne State University, Detroit, Michigan, United States\\
$^{139}$ Yale University, New Haven, Connecticut, United States\\
$^{140}$ Yonsei University, Seoul, Republic of Korea\\
$^{141}$  Zentrum  f\"{u}r Technologie und Transfer (ZTT), Worms, Germany\\
$^{142}$ Affiliated with an institute covered by a cooperation agreement with CERN\\
$^{143}$ Affiliated with an international laboratory covered by a cooperation agreement with CERN.\\

\end{flushleft} 
  
\end{document}